\documentclass[a4paper,12pt]{article}

\usepackage[section]{placeins} % float not across section
\usepackage{float} % not float with [H]
\usepackage{amsmath}
\usepackage{amsfonts}
\usepackage{graphicx}
\usepackage[colorinlistoftodos]{todonotes}
\usepackage{verbatim}
\usepackage{enumitem}
\usepackage[margin=1.0in]{geometry}
\usepackage{bbm}
\usepackage[super]{nth}
\usepackage{accents}
\usepackage{bookmark}
\usepackage{pdfsync} %!!!syc

\usepackage{amsmath}
\numberwithin{equation}{section}
\usepackage{booktabs}

\usepackage{subfigure} 

\usepackage{threeparttable}

\usepackage{parskip}

\usepackage{indentfirst}
\newcommand{\papertitle}{Short-term shock, long-lasting payment:\\
 Evidence from the Lushan Earthquake}
\newcommand{\yourname}{Yujue Wang}
\newcommand{\youremail}{ywang@wisc.edu}

\usepackage{natbib}
\setcitestyle{authoryear,open={(},close={)}}

\begin{document}

\begin{center}{\LARGE \scshape \papertitle}\end{center}

\begin{center}\vspace{0.2em} {\Large \yourname \\}{\youremail}\end{center}
\begin{center}{Dec 15, 2022}\end{center}

\vspace{50pt}

\begin{center}
\large{\textbf{Abstract}}
\end{center}

Abrupt catastrophic events bring business risks into firms. The paper introduces the Great Lushan Earthquake in 2013 in China as an unexpected shock to explore the causal effects on public firms in both the long and short term. DID-PSM methods are conducted to examine the robustness of causal inference. The identifications and estimations indicate that catastrophic shock significantly negatively impacts cash flow liquidity and profitability in the short term. Besides, the practical influences on firms' manufacturing and operation emerge in the treated group. Firms increase non-business expenditures and retained earnings as a financial measure to resist series risk during the shock period. As the long-term payment, the decline in production factors, particularly in employment level and the loss in fixed assets, are permanent. The earthquake's comprehensive interactions are also reflected. The recovery from the disaster would benefit the companies by raising the growth rate of R\&D and enhancing competitiveness through increasing market share, though these effects are temporary. PSM-DID and event study methods are implemented to investigate the general effects of specific strong earthquakes on local public firms nationwide. Consistent with the Lushan Earthquake, the ratio of cash flow to sales dropped drastically and recovered in 3 subsequent semesters. The shock on sales was transitory, only in the current semester.

\thispagestyle{empty}       
\newpage                   

\setcounter{page}{1}

\section{Introduction}

Sudden natural disasters introduce operational hazards to firms. Earthquakes are frequent natural disasters, strong and quick, with clear event time. Rather than meteorological disasters like hurricanes and floods, earthquakes are more unexpected and would not last for a long period. Furthermore, great destructive earthquakes usually cause numerous asset losses for firms, bring threats to people, and trigger panic in the whole economy.

Although earthquakes are sudden events, the risk propagation on the economy is long-lasting through multiple channels. Earthquakes' impact on firms' stock price and total assets is not limited to the short-term, nor in the directly covered areas (\cite{valizadeh2017ripple}). From the ex-post perspective, firms with factories located in areas affected by natural disasters are much less profitable(\cite{hsu2018natural}). In another bizarre way, earthquake shock could also spur the rise of innovation ability (\cite{rao2021}). In summary, earthquakes have comprehensive effects on firm operations and economic development.

\cite{cainelli2018natural} estimates the effect of the earthquake on firms' performance by Difference-in-differences and propensity score matching in levels and first-differences with data in Italian. Besides, a sales decrease is demonstrated, possibly brought by the blackout due to the Hokkaido earthquake \citep{ozaki2019estimation}. Event study(\cite{dou2022competition}) could be applied to the effect of strong natural disasters that happened at different times on firms in competition networks. Supply networks' vulnerability could be the fragile part to be broken when interrupted by natural shocks, so firms' supply network structures change after experiencing the catastrophic supply chain disruptions caused by the evidence of the 2011 Tohoku earthquake and tsunami in Japan \citep{son2021catastrophic}. (\cite{hyun2019spillovers}) illustrated the risk propagation mechanism and the spillover effects on firm networks. (\cite{chen2021regulating}) also attempts to add variables such as node degree and network topology centrality.

The second part of this paper will introduce the data sources and processing of earthquake data and firms' financial statements. The third part introduces the benchmark model with the application of DID. The fourth part introduces the identification of channels to investigate the real impact of the great Lushan Earthquakes on production, employment, and market ranking. The fifth part provides robust tests with the placebo test and PSM estimation to support the analysis. The sixth part implemented the PSM-DID based on previous analysis to examine hundreds of strong earthquakes' effects on local firms. The seventh part serves as a shred of strong evidence about the significance of such generalized issues, combined with the event study model. The eighth part includes a summary and suggestions.

\section{Data}
  \subsection{Earthquakes Frequency}
  According to data scraping from China Earthquake Administration (https://www.cea.gov.cn), from 2001 to 2021, there were 8115 earthquakes documented, measured by the Richter magnitude scale. The data contains each earthquake's event time, location, magnitude, and depth. More information refers to table \ref{eq1} on page \pageref{eq1}. Earthquakes larger than 4.5 magnitudes could be noticeable and felt by people. If magnitudes are larger than 6, earthquakes are called strong earthquakes and damage a moderate number of structures in populated areas. After the Great Wenchuan Earthquake in 2008, China attempted to establish a complete monitoring system. Therefore earthquake statistics after 2008 sustained a stable scale.

        \begin{table}
          \centering
          \scalebox{0.8}{
        \begin{threeparttable}         
        \caption{Earthquakes Frequency in China}
        \begin{tabular}{@{}cccccccccccccccc@{}}
        \toprule
        Year       & Before\tnote{1}  & 2008 & 2009 & 2010 & 2011 & 2012 & 2013 & 2014 & 2015 & 2016 & 2017 & 2018 & 2019 & 2020 & 2021 \\ \midrule
        Total      & 99     & 281  & 714  & 510  & 402  & 471  & 767  & 717  & 585  & 549  & 529  & 561  & 664  & 577  & 689  \\
        M$\geq$4.5 & 94     & 131  & 85   & 61   & 45   & 49   & 90   & 47   & 33   & 40   & 39   & 46   & 49   & 42   & 60   \\
        M$\geq$6   & 15     & 15   & 4    & 3    & 2    & 3    & 3    & 5    & 1    & 5    & 3    & 0    & 2    & 3    & 4    \\ \bottomrule
        \label{eq1}
        \end{tabular}
        \begin{tablenotes}  
        \footnotesize             
        \item[1] Aggregate earthquakes from 2001-2007.
        \end{tablenotes}           
        \end{threeparttable}}       
        \end{table}

  \subsection{Public firms}
  This paper selects firm-level annual and semester public information of firms listed in the Chinese stocks market from the China Stock Market and Accounting Research(CSMAR) database. Data mainly contains firms' annual reported information and the annual financial reports of each firm. The data ranges from 2008 to 2021. I chose 2008 as the starting point because earthquake statistics are stable at a hundred-scale per year starting from 2008. The financial report regulation for public firms in China experienced a reformation in 2007. Database describing the shareholding structure are merged to specify if the firm is controlled by the government and the capital stock in circulation share of the total capital stock.

  Stock symbols connect the above data. Particularly, firms with primary businesses in the finance field are dropped by the industry code. Null values are removed during filtering. All continuous variables are winsorized at the 1st and 99th percentiles. Variables from the balance sheet are standardized in millions and billions of Chinese Yuan for convenience.

\section{Benchmark regression and estimation}
\subsection{Regression model}
First, a single strong earthquake would intuitively result in operation distress for the firms. The Strong Lushan Earthquake in 2013 was chosen because it is one of the most severe earthquakes in the past two decades, ranked only second to the Great Wenchuan Earthquake in 2008. The magnitude of the Lushan Earthquake is 7.0. The direct economic loss consulted in Sichuan Province exceeds 130 billion dollars. Wenchuan Earthquake and Lushan Earthquake both happened in Sichuan province because Sichuan province and Chongqing are located geographically in earthquake-prone areas, i.e., Longmenshan Fault. 

Thus, the treatment group is listed firms in Sichuan, where the earthquake happened. Considering the similarity in the industry composition of public firms, Henan Province and Hebei Province are combined to serve as the control group. The control group firms are distributed mostly in the manufacturing and production industry. Refers to Table \ref{indus} on page \pageref{indus}.

\begin{table}[H]
  \centering
  \scalebox{0.8}{
  \begin{threeparttable}
  \caption{Industry Distribution}
\label{indus}
\begin{tabular}{llll}
\toprule
Industry & Freq in Treated & Freq in Control  \\ \hline
Agriculture, forestry, animal husbandry, fishing      & 7                     & 21                    \\
Mining     & 16                    & 70                    \\
Manufacturing     & 879                   & 1,251                 \\
Electricity, heat, gas and water production and supply industries     & 105                   & 67                    \\
 Contruction     & 47                    & 4                     \\
Wholesales and retail trade    & 53                    & 28                    \\
Transportation, warehousing and postal services     & 30                    & 35                    \\
Information transmission, software and information technology services     & 87                    & 43                    \\
Real estate    & 27                    & 29                    \\
Leasing and business service     & 10                    & 0                     \\
Scientific research and technical services   & 14                    & 8                     \\
Water conservancies, environment and public facility management    & 13                    & 5                     \\
Health and social work    & 1                     & 0                     \\
Culture, Sports and Entertainment    & 16                    & 11                    \\
Others   & 6                     & 9                    \\
 \bottomrule
\end{tabular}
\end{threeparttable}  }     
\end{table}

Besides, Henan and Hebei are also located in the Taihang earthquake zone, where the Strong Tangshan Earthquake (magnitude 7.8) happened in 1976, which supports the comparability in the geographical locations of the seismic zone. The model is constructed as below:

\begin{equation}
  Influenced\_items_{i,t}/Firm\_scale_{i,t}=\alpha+X'_{i,t}\Gamma+ \tau post_{i,t}*Treat_{i,t}+\theta_i+\delta_t+ indus_i+\epsilon_{i,t}
\end{equation}

  $Influenced\ items$ are general terms that could be highly correlated to the sudden fluctuations caused by earthquakes. Specifically, $Influenced_items$ refers to the cash flow of each public firm in each year. $Firm\_scale$ is measured by sales and assets to derive the dependent variable as a ratio. The controls $X_{st} $  include  $ CFA,GROW$ $ , LRL,ROA,ROE,$ $ eq\_times$  and so on in each year for each firm. The variables applied in this paper are listed in the \ref{var} on page \pageref{var}. $ROE$ is the return on equity. $ROA$ is the return on assets, reflecting the profitability. $LRL$ is the ratio of net profits and sales, i.e., the profit margin of the primary business. Eq times is the moderate and strong earthquake times (magnitude is higher than 4.5) of the city area where the firm's office is located, merged from the earthquake database established before by the administrative code of China (https://www.mca.gov.cn/article/sj/xzqh/1980/). 

  $\epsilon $ is a stochastic error. $\theta_i $ is firm-level fixed effect, $\delta_t$ is year fixed effect, and $indus\_{i,t}$ is the industry fixed effects. $post_{i,t}$  takes the value of 1 after the earthquakes happen, i.e., including 2013 and after. Parameter $\tau$ captures the average impact of unexpected natural disasters. All statistics are annual.

  \begin{table}[H]
    \centering
    \caption{Variable Meanings}
    \scalebox{0.6}{
    \begin{threeparttable}         
    \begin{tabular}{llll}
      \textbf{Original Variables} \\
      \toprule
             Name & Variable Name   & Meaning     & Sourse      \\ \hline
             Cash Flow & cf   & The difference between cash inflow and cash outflow     & Cash Flow Statement     \\
             Sales & sales &  The sum of all revenues generated & Income Statement\\
             R\&D Spenditure & rd &  The sum of spenditure relative to R\&D & Income Statement\\
             Expenditure beyond business&non bz expend&Expenditure not on the main business costs&Income Statement\\
             Net Profit & net profit  &  Net profit & Income Statement\\
             Total Assets &  ttl assets & Total assets  &Balance Sheet\\
             Fix Assets &  fix assets net & Total fixed assets  &Balance Sheet\\
             Intangible  Assets & intangible assets net & Total intangible assets  &Balance Sheet\\
             Current Asset & total current assets & Total current assets  &Balance Sheet\\
             Net Assets  & net assets & Total of shareholders' equity items &Balance Sheet\\
             Total Liability  & total liab & Total liabilities &Balance Sheet\\
             Current Liability  & current liab & Total current liabilities &Balance Sheet\\
             Reserve Fund & reserve& Reserve fund drawn from profits &Balance Sheet\\
             Undistribution Profits & undis & Undistribution profits  &Balance Sheet\\
            Actual Controller & gov & Whether controlled by governments & Annual Reports\\
            Circulation Stock& nshr nocrt & Capital stock in circulation & Annual Reports\\
            Total Stock& nshrttl & Capital stock in total & Annual Reports\\
            List Age& lsitage & Listed year until 2022 & Annual Reports\\ 
           Separation Degree and &separation& Separation ratio of ownership and control & Annual Reports\\         
            Age& age & Established year until 2022 & Annual Reports\\
            Employee& Employee &Total employees& Annual Reports\\
            Top10 Holders Rate& toptenholdersrate &The sum of stockholding rate  of top10 holders& Annual Reports\\
            Earthquake Times&eq times & Annual earthquake times in the district the office is located &Web Scraping data \\
    \vspace{4pt}\\
\textbf{Generated Variables}  \\ \hline 
Tangible Assets &  tangi & Sum of  fix assets and total current assets   &fix assets net+ttl current assets \\           
             Cash Flow to Sales &  cfs & The ratio of cash flow to sales     & cf/sales     \\
             Cash Flow to Assets &  cfa & The ratio of cash flow to assets     & cf/ttl assests     \\
             Cash Flow to Fixed Assets &  cffa & The ratio of cash flow to fixed assets     & cf/fix assests net     \\
             Cash Flow to Tangible Assets &  cfta & The ratio of cash flow to tangible assets     & cf/tangi     \\
             Retained Earnings&retained earnings & Sum of reserve fund and undistribution profits to resist risks & reserve+undis \\
             Grow & grow& The growth potential of firms & retained earnings/net assets\\
             Profit Margin & lrl & Profit margin from primary business&net profit/sales\\
             Leverage& lev & Leverage ratio&total liab/ttl assets\\
              Profitability & roa& The capability of profitability & net profit/ttl assets\\
              Profitability by Net Assets & roe& The capability of profitability & net profit/net assets\\
              Sales Growth&sg &Sales growth rate &log(sale/L.sale)\\
              Profit Growth&pg &Profits growth rate &log(net profit/L.net profit)\\
              R\&D Growth&rdg & R\&D growth rate &log(F.rd/rd)\\
              Total Sales in Subdivided Industry& salesvalue & Sum of sales of all firms in a specific industry & $\sum $ sales\\
              Loss& loss &The motive to avoid loss &1(net profit<0)\\
              Non-circulation to Total  & non-tradable shares &The ratio of non-tradable stock shares to the total  &nshr nocrt/nshrttl \\
              Log Total Assets&lassets &Natural logarithm of total assets &log(ttl assets)\\
              Log Net Assets&lnnass &Natural logarithm of net assets &log(net assets)\\
              Log Total Liability&lliab &Natural logarithm of total liability &log(total liab)\\
              Log Cash Flow&lcf &Natural logarithm of cash flow  &log(cf)\\
              Log Tangible Assets&ltangi &Natural logarithm of tangible assets  &log(tangi)\\
              Log Salesvalue&lsalesvalue &Natural logarithm of salesvalue &log(salesvalue)\\
              Relative Market Share&rsms &Sales divided by the largest sales in the same industry&sales/max(sales)\\
              Log Labor&llabor &Natural logarithm of Employee &log(Employee)\\ \bottomrule
    \end{tabular}
    \label{var}
  \end{threeparttable} } 
    \end{table}

  \subsection{DID on Lushan Earthquake}

  To examine the average trade effect of sudden earthquake on the firms in Sichuan, here regress on the following model:

\begin{equation}
  cf_{i,t}/sales_{i,t}=\alpha+X'_{i,t}\Gamma+ \tau post_{i,t}*Treat_{i,t}+\theta_i+\delta_t+\epsilon_{i,t}
\end{equation}
  Table \ref{sum} on page \pageref{sum} and Table \ref{corrmat} on page \pageref{corrmat} demonstrates the distribution of main control varibles and correlations among them. The correlation matrix is strong enough to rule out the probability of multi-colliearity, which could results in biased estimation on the coefficients of ATT.\\

        \begin{table}[H]
        \centering
        \caption{Summary of Statistics}
        \scalebox{1}{
        \begin{threeparttable}         
        \begin{tabular}{lllllllll}
          \toprule
          \textbf{Treat: 0}  \\ \hline
          & N    & mean     & sd       & min        & p25     & Median   & p75      & max       \\
          cfs           & 1578 & .0953    & .1553    & -.4807     & .0171   & .078     & .1667    & .6114     \\
          cfa           & 1578 & .0505    & .0706    & -.1625     & .0093   & .0453    & .088     & .252      \\
          cf & 1578 & 668.1248 & 1622.383 & -1660.8157 & 19.1727 & 130.4805 & 605.1819 & 9434.5336 \\
          grow          & 1578 & .1242    & 1.1213   & -9.8156    & .1779   & .3033    & .4531    & 2.0211    \\
          lrl           & 1578 & .0709    & .1812    & -1.1326    & .0206   & .0677    & .1391    & .6229     \\
          roa           & 1578 & .0405    & .0656    & -.2621     & .0115   & .0377    & .07      & .2284     \\
          roe           & 1578 & .0618    & .1821    & -1.2086    & .0274   & .0728    & .1212    & .5937     \\
          lassets       & 1578 & 8.3708   & 1.4703   & 5.2651     & 7.3317  & 8.2552   & 9.4191   & 11.9242 \\ \hline 
\vspace{4pt}\\
\textbf{Treat: 1}  \\ \hline

cfs           & 1308 & .1073    & .1776     & -.4807     & .0173   & .091     & .1959    & .6114     \\
cfa           & 1308 & .0485    & .0686     & -.1625     & .0085   & .0476    & .087     & .252      \\
cf & 1308 & 433.7782 & 1190.4247 & -1660.8157 & 15.1178 & 110.6426 & 339.9948 & 9434.5336 \\
grow          & 1308 & .0701    & 1.3739    & -9.8156    & .1518   & .3124    & .4572    & 2.0211    \\
lrl           & 1308 & .0789    & .2261     & -1.1326    & .0264   & .0784    & .1642    & .6229     \\
roa           & 1308 & .0402    & .0709     & -.2621     & .0144   & .039     & .0734    & .2284     \\
roe           & 1308 & .062     & .2041     & -1.2086    & .0329   & .0747    & .1237    & .5937     \\
lassets       & 1308 & 8.0305   & 1.3428    & 5.2651     & 7.0311  & 7.9037   & 8.7652   & 11.9242   \\ \bottomrule

        \end{tabular}
        \label{sum}
        \end{threeparttable}     } 
        \end{table}

            \begin{table}[H]
              \centering
              \begin{threeparttable}
              \caption{Pairwise correlations matrix of variables}
            \label{corrmat}
            \begin{tabular}{lllllllll}
            \toprule
            Variables   & (1)     & (2)     & (3)     & (4)     & (5)     & (6)     & (7)     & (8)   \\ \hline
            (1) cfs     & 1.000   &       &         &         &         &         &         &       \\
            &         &         &         &         &         &         &         &       \\
(2) cfa     & 0.788*  & 1.000   &         &         &         &         &         &       \\
            & (0.000) &         &         &         &         &         &         &       \\
(3) lnnass  & 0.177*  & 0.145*  & 1.000   &         &         &         &         &       \\
            & (0.000) & (0.000) &         &         &         &         &         &       \\
(4) grow    & 0.181*  & 0.180*  & 0.389*  & 1.000   &         &         &         &       \\
            & (0.000) & (0.000) & (0.000) &         &         &         &         &       \\
(5) lassets & 0.108*  & 0.083*  & 0.930*  & 0.158*  & 1.000   &         &         &       \\
            & (0.000) & (0.000) & (0.000) & (0.000) &         &         &         &       \\
(6) lrl     & 0.373*  & 0.285*  & 0.097*  & 0.172*  & 0.010   & 1.000   &         &       \\
            & (0.000) & (0.000) & (0.000) & (0.000) & (0.589) &         &         &       \\
(7) roa     & 0.325*  & 0.430*  & 0.095*  & 0.230*  & -0.024  & 0.826*  & 1.000   &       \\
            & (0.000) & (0.000) & (0.000) & (0.000) & (0.194) & (0.000) &         &       \\
(8) roe     & 0.205*  & 0.284*  & 0.152*  & 0.345*  & 0.023   & 0.519*  & 0.621*  & 1.000 \\
            & (0.000) & (0.000) & (0.000) & (0.000) & (0.219) & (0.000) & (0.000) &      \\
            \bottomrule
            \end{tabular}
            \begin{tablenotes}
            \footnotesize
            \item[]
             $^*p < .1$ \\
            \end{tablenotes}            
            \end{threeparttable}      
            \end{table}

\begin{figure}
  \caption{Parallel Trend on dependent variable}
  \centering
  \includegraphics[width=10cm]{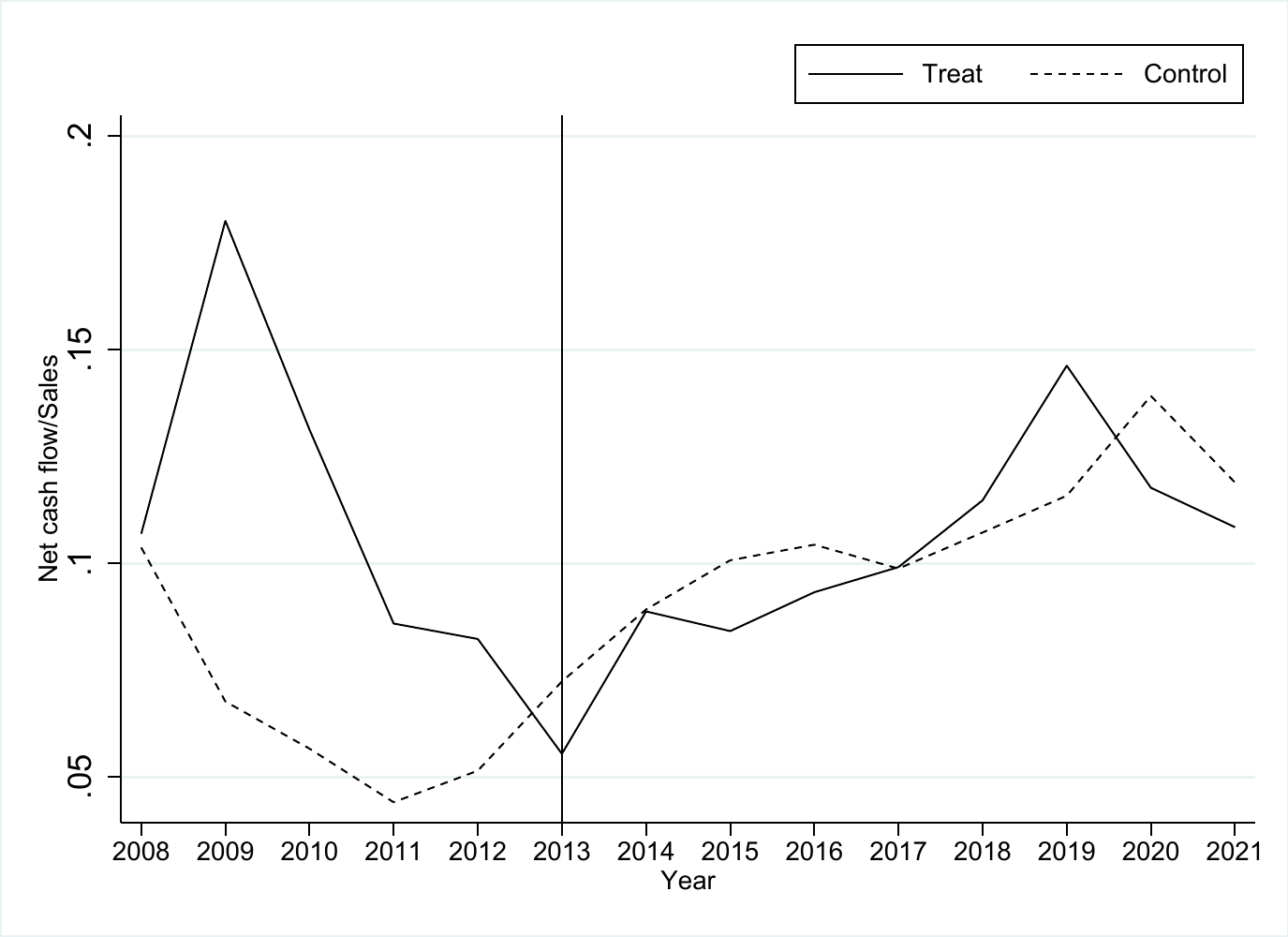} 
  \label{pt2}
  \end{figure}

            According to Figure \ref{pt2} on page \pageref{pt2}, the dependent variable preliminarily has a similar trend before the earthquake. In 2013, the ratio experienced a sharp decrease, which generated an obvious cross between two curves around 2013. The trend before and after 2 years of the shock time is almost parallel. The unusual boost in 2009 of the treatment group is most likely to benefit from recovery from the Wenchuan Great Earthquake, the most severe earthquake in the past three decades. Thus, in 2009, production and reconstruction reached a high level after the disaster under support from governments and social organizations. The peak in 2009 has a similar shape to the peak in 2014, which illustrates the reasonability of the analysis. After Covid-19, both treated and controlled firms suffer from a sluggish economy. We expected a short-term shock directly on the ratio, which is consistent with the results. Dynamic effects are necessary to demonstrate robustness on the average treatment effects.

\begin{table}[H]
  \centering
  \begin{threeparttable}
  \caption{Firm cash flow changes by the great earthquake}
\label{did0}
\begin{tabular}{lllll}
\toprule
& (1)      & (2)              & (3)        & (4)                               \\
& \multicolumn{4}{c}{$Net\  cash\  flow_{i,t}/Sales_{i,t} $}            \\ \cmidrule(l){2-5}

ATT               & -.021869*** & -.023182**   & -.022339*** &     -.021938***         \\
                  & (.00788)    & (.008195)    & (.008032)   &      (.008101)       \\
Firm FE           & No          & NO          & Yes         & Yes         \\
Year FE           & No          & Yes         & No          & Yes         \\  \hline
cfa               & 1.945973*** & 1.981649***  & 1.923844*** & 1.925862*** \\
                  & (.062813)   & (.052093)    & (.065424)   & (.065113)   \\
grow              & .005972**   & .007749**    & .004588*    & .004527*    \\
                  & (.002435)   & (.00269)     & (.002404)   & (.002451)   \\
lrl               & .241514***  & .42131***    & .158216***  & .157443***  \\
                  & (.043378)   & (.03789)     & (.038092)   & (.038606)   \\
roa               & -.676206*** & -1.099858*** & -.496688*** & -.498329*** \\
                  & (.115777)   & (.097895)    & (.108399)   & (.10946)    \\
roe               & -.025952*   & -.037451*    & -.018699    & -.018387    \\
                  & (.014944)   & (.020708)    & (.014615)   & (.014379)   \\
lassets           & .005868**   & .00087       & .013498***  & .014775***  \\
                  & (.002951)   & (.001048)    & (.004346)   & (.005335)   \\
gov               & -.005809    & .003938      & -.017838    & -.016345    \\
                  & (.007318)   & (.003911)    & (.011403)   & (.01142)    \\
cons             & -.047354**  & .002438      & -.088347**  & -.093191**  \\
                  & (.021989)   & (.007863)    & (.034993)   & (.045644)   \\
Observations      & 2886        & 2886         & 2886        & 2858        \\
R-squared         & 0.7124           & .708545      & .717161     & .849205    \\
\bottomrule
\end{tabular}
\begin{tablenotes}
\footnotesize
\item[]
Standard errors are in parentheses.\\
$ ^{***} p<.01, ^{**} p<.05, ^* p <.1$\\
\end{tablenotes}        
\end{threeparttable}       
\end{table}
According to the results in figure \ref{did0} on page \pageref{did0}, the shock negatively affects the cash flow standardized by sales. All four models are robust to heteroskedasticity. The model (4) is estimated by the Multi-Way Fixed Effects method to overcome slow convergence. Comparing the four cases by whether fixed effects of firms and years are added, the model with firm fixed effects and year effects has the highest within R-Squared. The magnitude of ATT in model(4) is slightly less than that in the model (1). Among all control variables, the ratio of cash flow to total assets, profit margin, and profitability capability provides consistent superiority as control variables.

\begin{figure}
    \caption{Parallel Trend on coefficients}
    \centering
    \includegraphics[width=10cm]{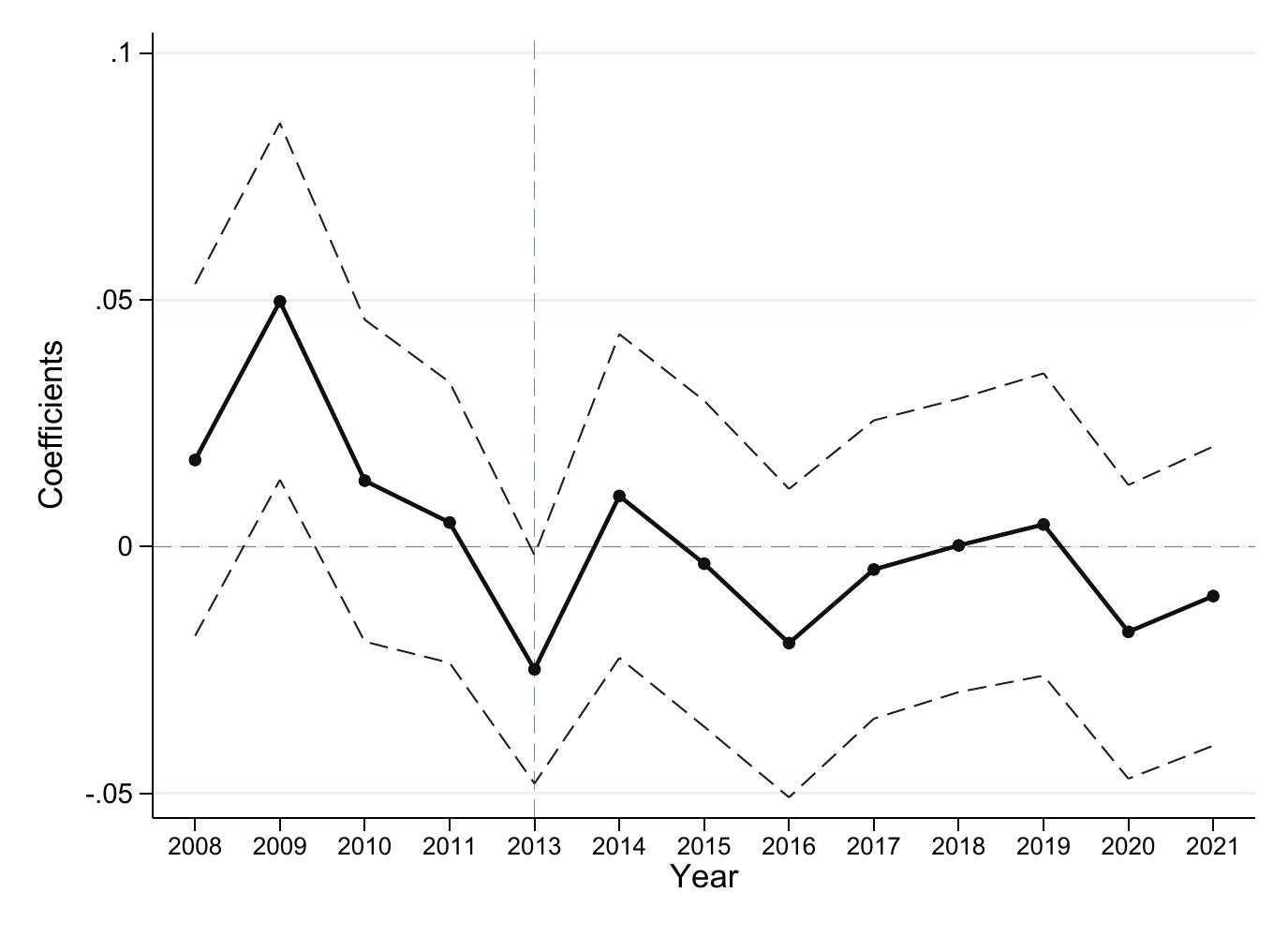} 
    \label{pt1} 
    \end{figure}

The coefficients of dynamic effects are demonstrated in Figure \ref{pt1} on page \pageref{pt1}. The area between dash curves is 95\% significance level of the estimated coefficients, which is the product of the year dummy variable and treat variable. 2012 (t=-1) is selected as the base year, i.e.the horizontal axis of y=0. The coefficient is only significantly negative in the year of the earthquake, which is consistent with our expectation of the short-term strike on cash flows. Cash flow is less likely to drop permanently. To sum up, the parallel trend test is robust enough to imply earthquakes' negative causal effects on firms' scaled cash flows. 

\section{Identification of Channels and Mechanisms}

  \subsection{Identification on the most direct effects}
  \begin{table}[H]
    \centering
    \scalebox{0.7}{
    \begin{threeparttable}
    \caption{Cash flow-Assets}
  \label{did2}
  \begin{tabular}{lllllllll}
  \toprule
   &(5)         & (6)        & (7)         & (8)        & (9)       & (10)         & (11)    &(12)    \\
        & cfs         & lsales     & cfa         & cffa       & cfta      & lcf         & ltangi      & roa     \\ \hline
  did          & -.056932*** & .01821     & -.017686*** & .548718    & -.023767**  & -.294972*** & -.023236    & -.014004**   \\
        & (.015725)   & (.053044)  & (.006395)   & (.587048)  & (.0112)    & (.106733)   & (.048165)   & (.005935)  \\
  grow         & .032672***  & .100991*   &             &            &       & & &      \\
        & (.012348)   & (.052858)  &             &            &          & & &    \\

        net profit   & & & & &  & .000072***  &             &             \\
        & & & & & & (.000024)   &             &             \\
  lassets      & .109893***  & .560295*** &             &            &      & & &        \\
        & (.036154)   & (.157934)  &             &            &     & & &         \\
  lnnass       & -.101515*** & .247269    &             &            &   & & &           \\
        & (.036205)   & (.152847)  &             &            &         & & &     \\
  lev          & -.262507*** & .805599**  &             &        & &    &   -.11712***           \\
        & (.082361)   & (.358635)  &             &            &   & &   & (.024475)        \\
  age          & .141061**   & .222741    & .064784***  & 2.029145   & .096715**   \\
        & (.056337)   & (.174283)  & (.021103)   & (1.745554) & (.038496)   \\
   listage & & & & &           &          &   & -.013951***\\
    & & & & &       &          &   & (.003598)\\    
  lsales       &             &            & .027375***  & -1.22985   & .033869*** & .831388***  & .394075***  & .037934*** \\
        &             &            & (.005208)   & (1.489263) & (.009599)  & (.053905)   & (.034778)   & (.00633) \\
  lliab        &             &            & -.01961***  & .880718    & -.019539***  &             & .369691***  & -.016476***\\
        &             &            & (.003987)   & (1.041002) & (.007124)    &             & (.034367)   & (.005958)  \\
  cons        & -.311093    & .059567    & -.194064*** & -2.749339  & -.316891***& -.982143**  & 2.330324*** & -.04325   \\
        & (.20653)    & (.57997)   & (.064925)   & (6.490113) & (.118676)  & (.405582)   & (.224538)   & (.029321)  \\
  Observations & 2821        & 2821       & 2858        & 2858       & 2858   & 2292        & 2858        & 2739      \\
  R-squared    & .483973     & .963326    & .442985     & .063591    & .405514  & .793138     & .975204     & .528773 \\
  \bottomrule
  \end{tabular}
  \begin{tablenotes}
  \footnotesize
  \item[]
  Standard errors are in parentheses.\\
  $ ^{***} p<.01, ^{**} p<.05, ^* p <.1$\\
  \end{tablenotes}           
  \end{threeparttable} }
  \end{table}
  
  Although the negative average treatment effects on the treated group are evident, it is still ambiguous about which variables are probably most directly shocked. The Table \ref{did2} on page \pageref{did2} is combined to identify the channel. (5)-(12) are estimated by Multi-Way Fixed Effects based on (4). Both firm fixed effects and year fixed effects are applied to ensure the estimation is robust to heteroskedasticity. Although the negative treated effects are significant to the ratio of cash flow to sales, the effects are not significant on the sales level. The ATT magnitude in (5) is -.056932 is higher than that in (4), i.e., .021938, because the control variables in (5) are adjusted based on multi-way fixed effects. The coefficients of the control variables also demonstrate that the estimation in (5) should be more precise than in (4).

  Comparison between (5), (6), and (10) indicates that cash flow violates most. Sales may not vary as drastically as cash flow because public firms' sales are comprehensive and relatively more stable than those of smaller firms. Given that the main composition of firms in the treated group and control group is manufacturing enterprise, the sustainable supply chain could support the firms to pull through the strike smoothly. The firms have a tendency to cut down the cash flow to resist the risk actively or cover the deficit passively.

  (7), (8), and (9) imply that the ratio of cash flow to total assets is more highly influenced than the ratio of cash flow to fixed effects and cash flow to tangible assets. The Combinition of (9) and (11) indicates that it is still the cash flow that is strong enough to manipulate the overall effects on the ratios. (5) and (7) are consistent because the log level of assets and sales are controlled respectively to examine the effects on cash flow to sales and assets.

  (7) and (12) mutually confirm that the profit has a similar trend with cash flows. According to (10), profits and cash flow have a positive correlation, while we cannot causally determine the relationship between profit and cash flows directly from the regressions here.

  \subsection{Real Impact}

  The real impact on firms is not as intuitive as the causal effect on profits and cash flows. Three dependent variables here are examined. A firm's absolute market share is its annual sales divided by the total sales in its industry. Nevertheless, the absolute market share is not comparable among a cohort where firms are distributed in different subdivided industries considering the distinguished concentration ratio of industries. Thus, relative market share, the firm's annual share divided by the maximal sales in the industry, is introduced as a standardized variable. The R\&D growth is calculated by dividing the expenditure in the last period.  

  According to (13) of Table \ref{did4} on page \pageref{did4}, the relative market share, indicating the relative market competitiveness, increased after the shock in 2013. The firms in the treated group, i.e., Sichuan province, had higher market competitiveness among all firms nationwide within the same industry. A higher level of capital (represented by cash flow and assets), and labors promote the relative market share of firms. The non-tradable stock capital refers to the stocks of listed companies in the Chinese stock market that cannot be freely traded on the trading market. The trade of such stocks could only be conducted by auctions, through agreements, or public trade to protect the rights of shareholders of other tradable shares. The ratio of non-circulation stock shares to the total stock shares indicates the financial liquidity of firms in the stock market, which is higher for those firms who are less able to seek financing support from the stock market to transfer financial risks. To sum up, the firms with less financing flexibility benefit less in the boost of relative market share after the earthquake. Besides, the larger scale of the subdivided industry also negatively acts on the rise of market position. The absolute market share varies less violently than the relative market share in magnitude, though they are not distinguished in the sign of coefficients. The regression is robust to heteroskedasticity by clustering in the year and firm level, containing the fixed effects in year, industry and firms as follows.

\begin{equation}
  Market Share=\alpha+X'_{i,t}\Gamma+ \tau post_{i,t}*Treat_{i,t}+\delta_t+\theta_i+ indus_i+\theta_i*indus_i +\epsilon_{i,t}
\end{equation}

The labor (15) was reduced after the earthquake prominently. The firms have the tendency to cut down employment to increase liquidity in cash flow and circulation funds. Particularly, for the firms controlled mainly by government institutions and having a higher equity concentration ratio, the labor reduction was more significant. Furthermore, higher return on assets prompts firms to cut down the less profitable production factor, like the labor force to maintain the profitability of assets. 

The growth rate of research and development (16) elevated significantly, which is consistent with the evidence from \cite{rao2021}. An outstanding sales growth rate would definitely promote the firm to spare more revenues to invest higher in technology. The profit last term would limit the available input in research. Moreover, extra expenditure beyond primary business also restricted the upper bound of such investment. 

(17) and (18) could be combined to illustrate how firms chose to report the loss caused by earthquakes to investors. Retained earnings are the sum of the reserve fund and undistributed profits to resist risks. Furthermore, any expenditure not out of manufacturing costs in primary business operations should be counted as a non-business expenditure. Firms responded to the shock of earthquakes in these two items rapidly. The earthquake increased the non-business expenditure significantly for the treated firms. The positive sign of the coefficients in (17) could be interpreted as the active preparation for earthquake risks.

Given that the majority composition of the firms in the two groups is manufacturing firms, we consider the direct fixed assets loss. As (19) shows, the loss resulting from an earthquake is significant enough with the firm scales controlled by sales and tangible assets.

\begin{figure}[H]
  \centering

  \subfigure[Relative Market Share (Base:2013. CI:90\%)]{
  \begin{minipage}[t]{0.5\linewidth}
  \centering
  \includegraphics[width=3in]{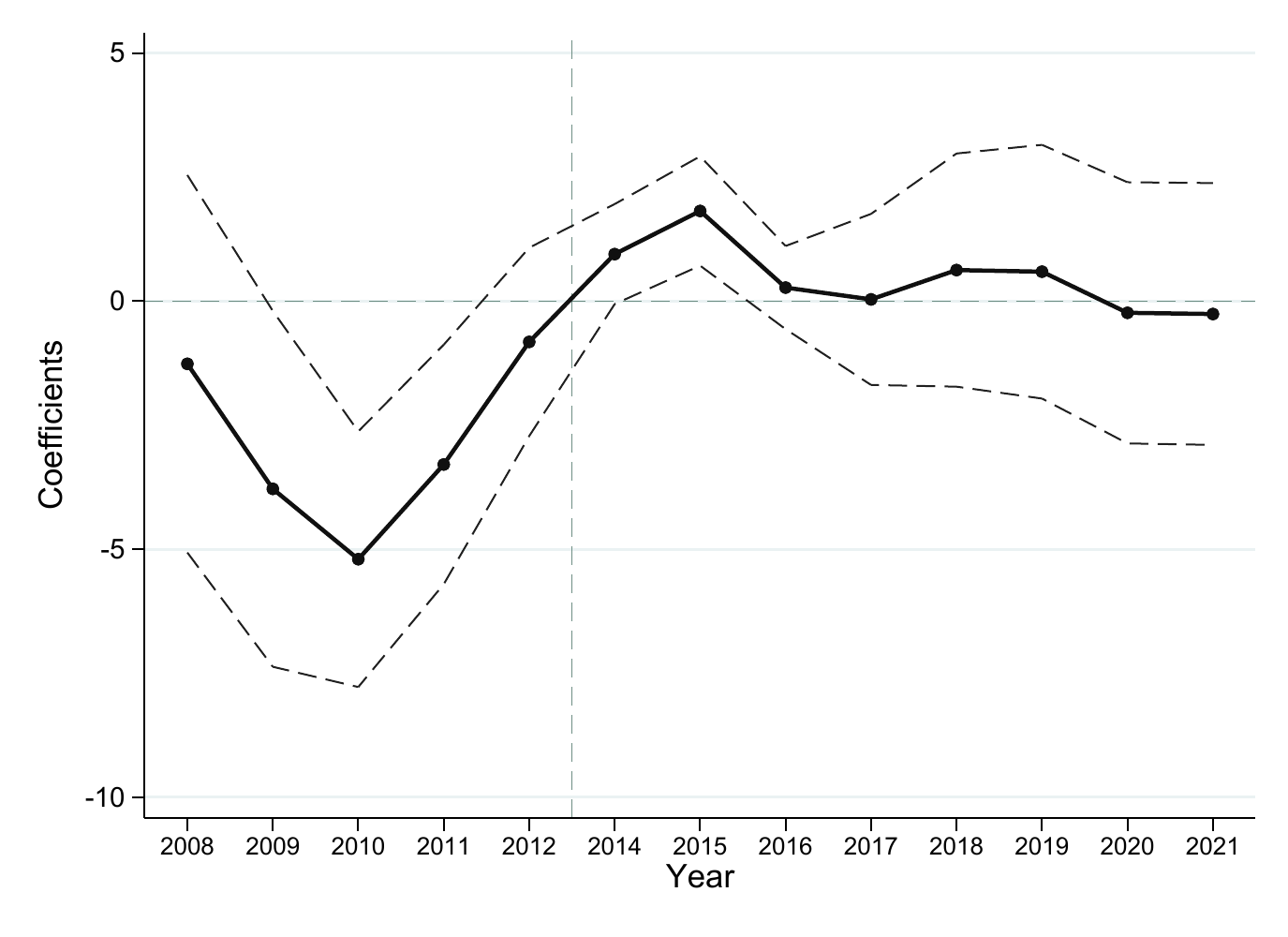}
  \end{minipage}%
  }%
  \subfigure[Absolute Market Share (Base:2013. CI:90\%)]{
  \begin{minipage}[t]{0.5\linewidth}
  \centering
  \includegraphics[width=3in]{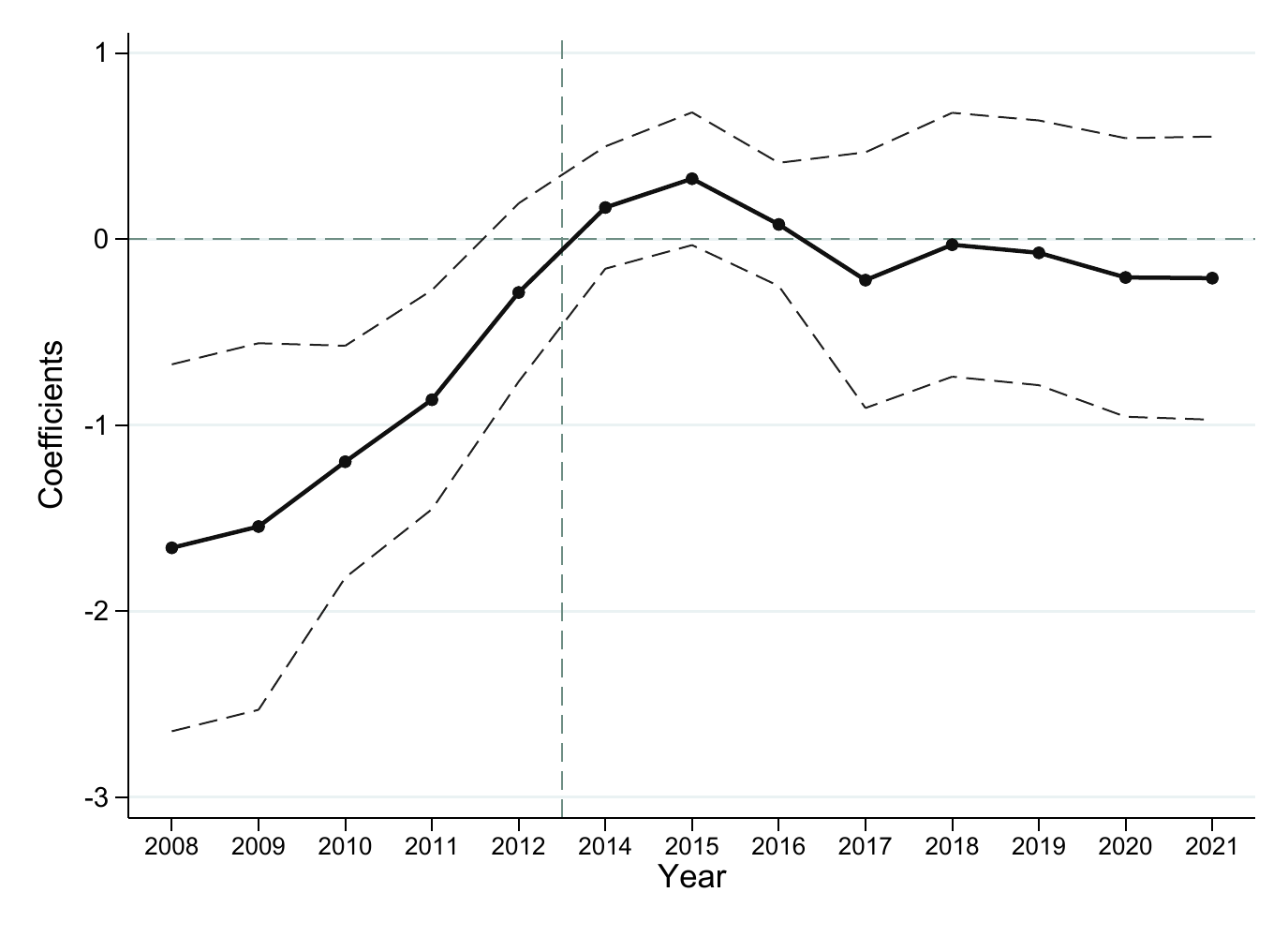}
  \end{minipage}%
  }%
        
  \subfigure[Labors (Base:2013. CI:90\%)]{
  \begin{minipage}[t]{0.5\linewidth}
  \centering
  \includegraphics[width=3in]{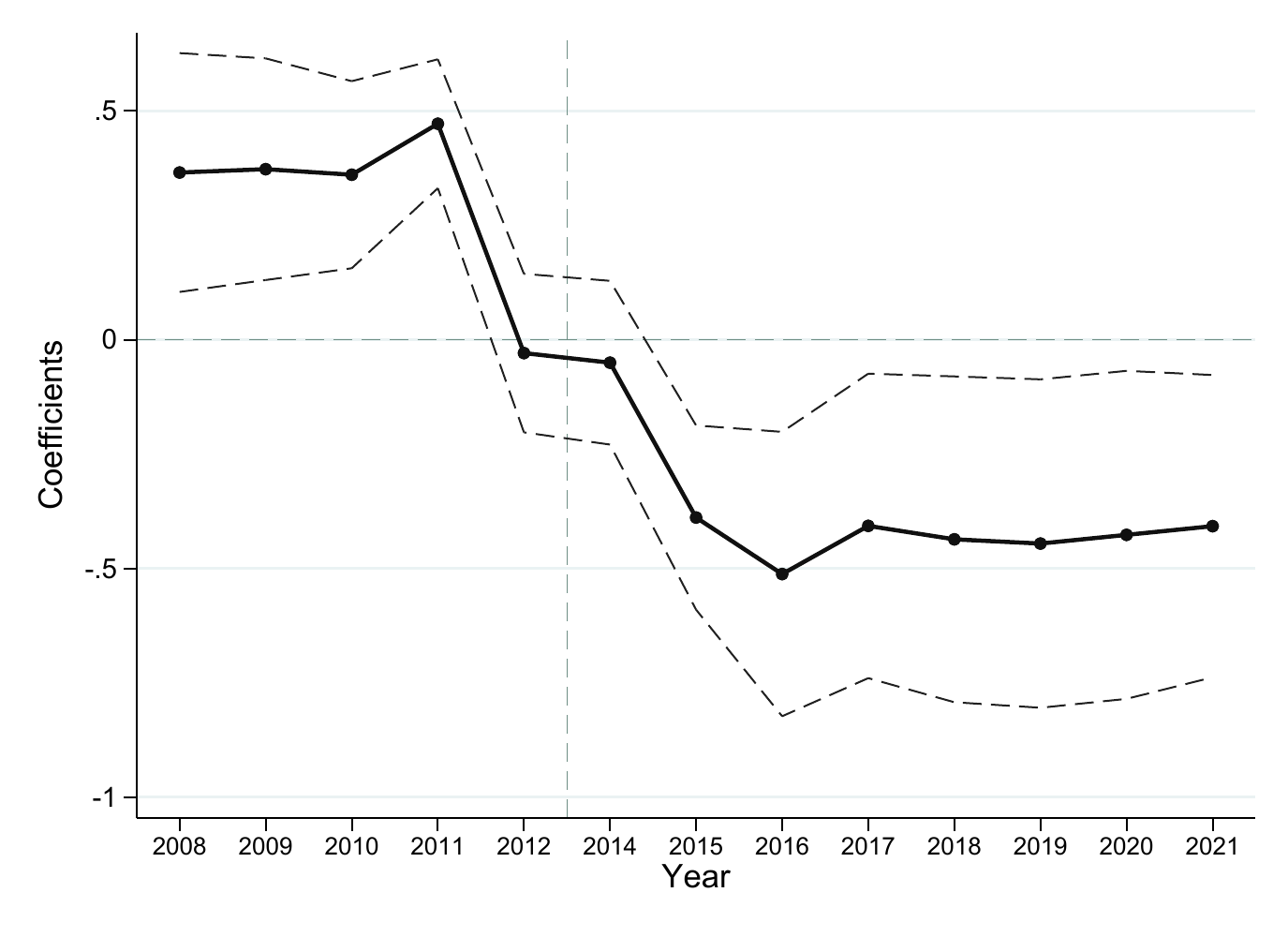}
  \end{minipage}%
  }%
  \subfigure[R\&D Growth Rate (Base:2013. CI:90\%)]{
  \begin{minipage}[t]{0.5\linewidth}
  \centering
  \includegraphics[width=3in]{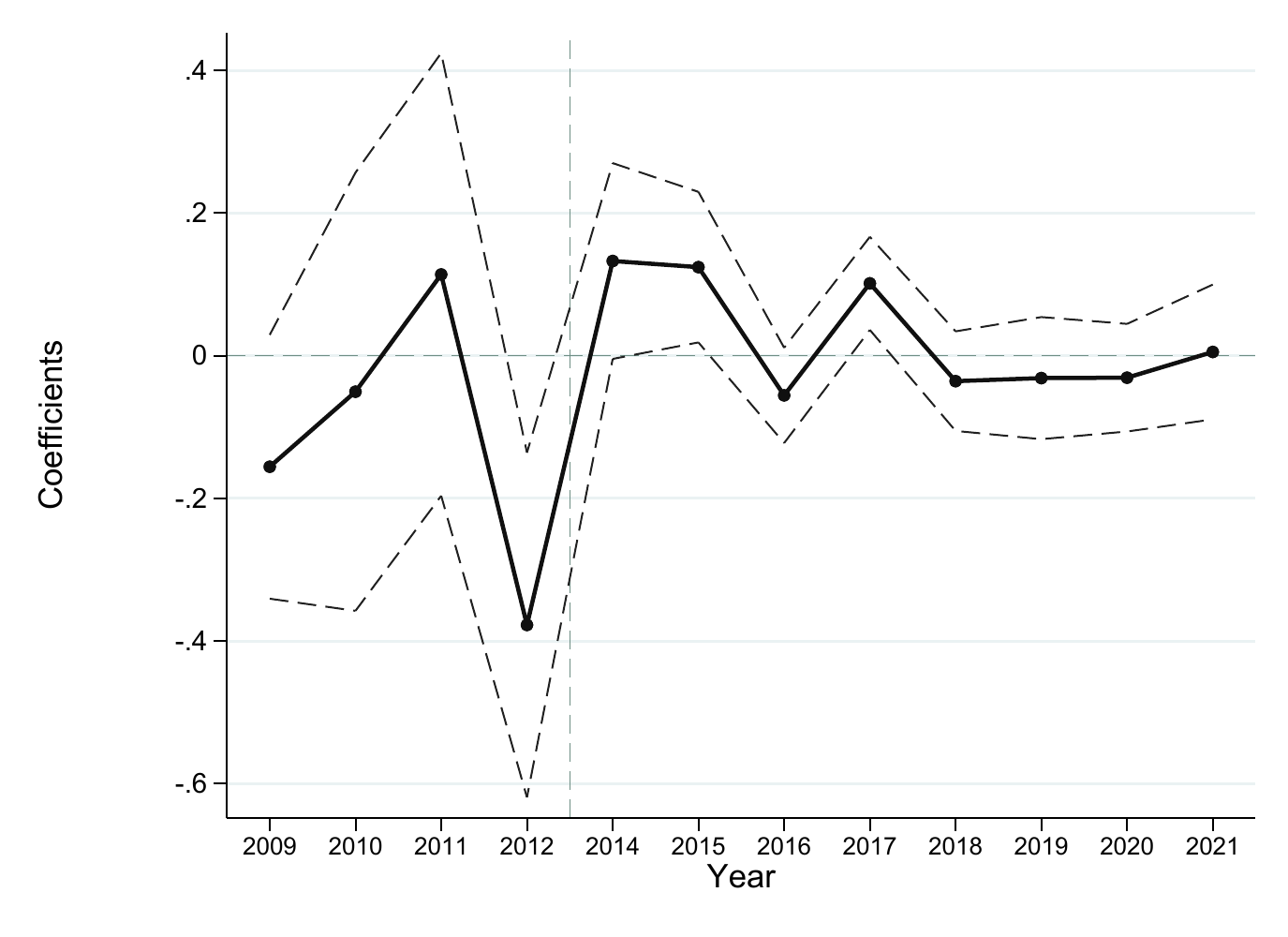}
  \end{minipage}% 
  }% 

  \subfigure[Retained Earnings (Base:2012. CI:90\%)]{
   \begin{minipage}[t]{0.33\linewidth}
   \centering
   \includegraphics[width=2in]{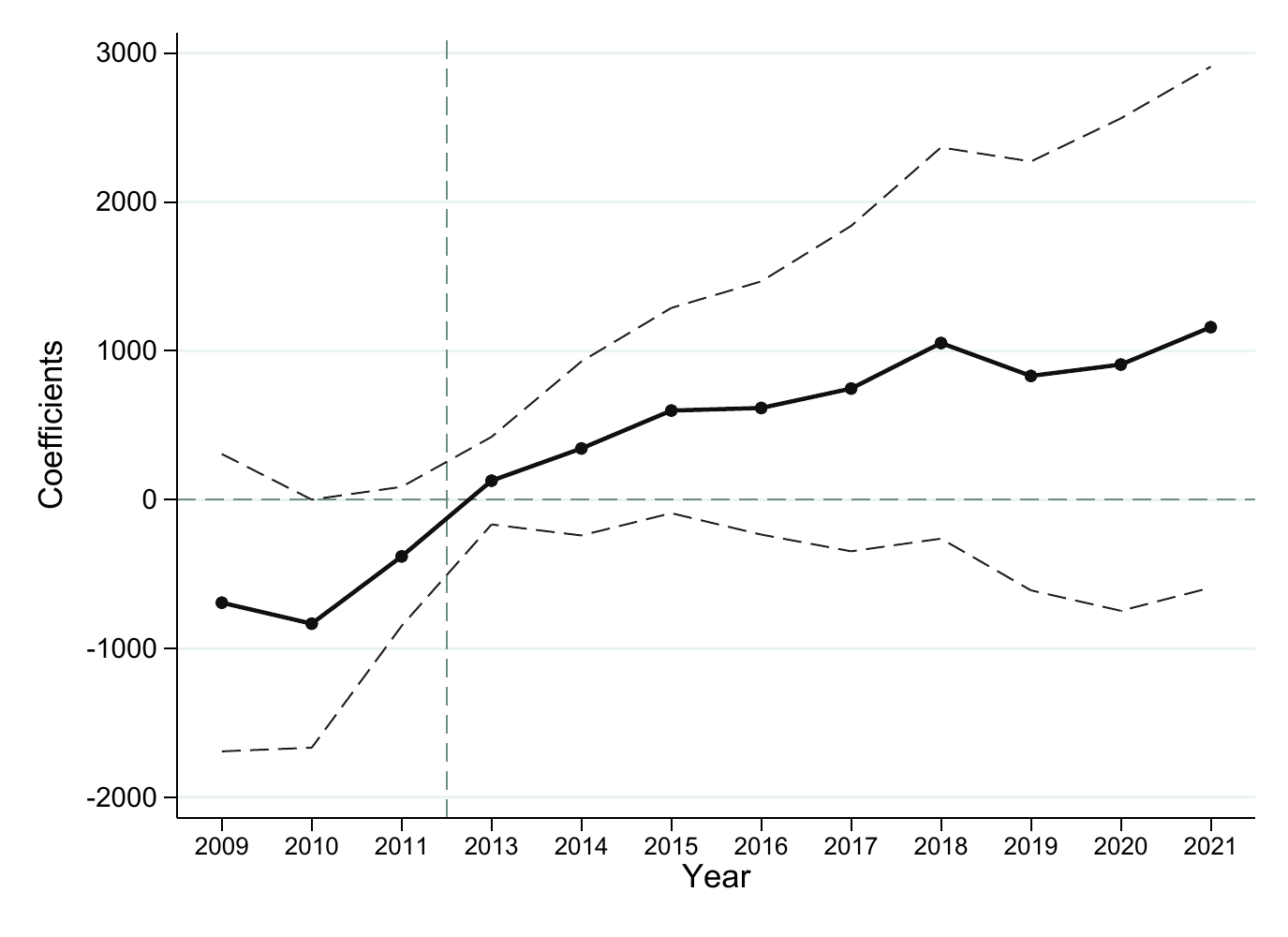}
   \end{minipage}%
      }%  
   \subfigure[Non-Biz Expenditure (Base:2012. CI:90\%)]{
   \begin{minipage}[t]{0.33\linewidth}
   \centering
   \includegraphics[width=2in]{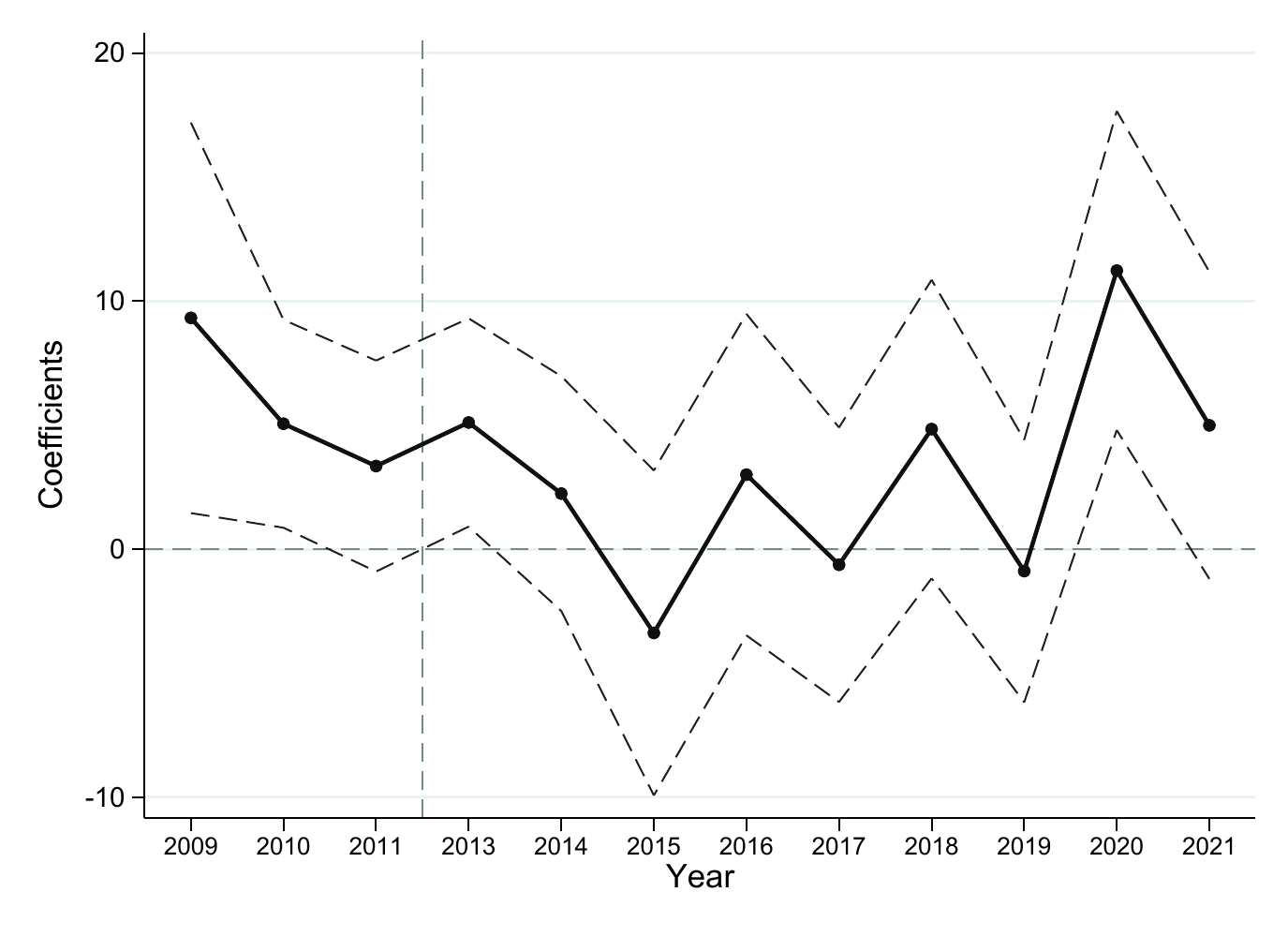}
   \end{minipage}%
       }%
   \subfigure[Fixed Assets (Base:2012. CI:90\%)]{
    \begin{minipage}[t]{0.33\linewidth}
    \centering
    \includegraphics[width=2in]{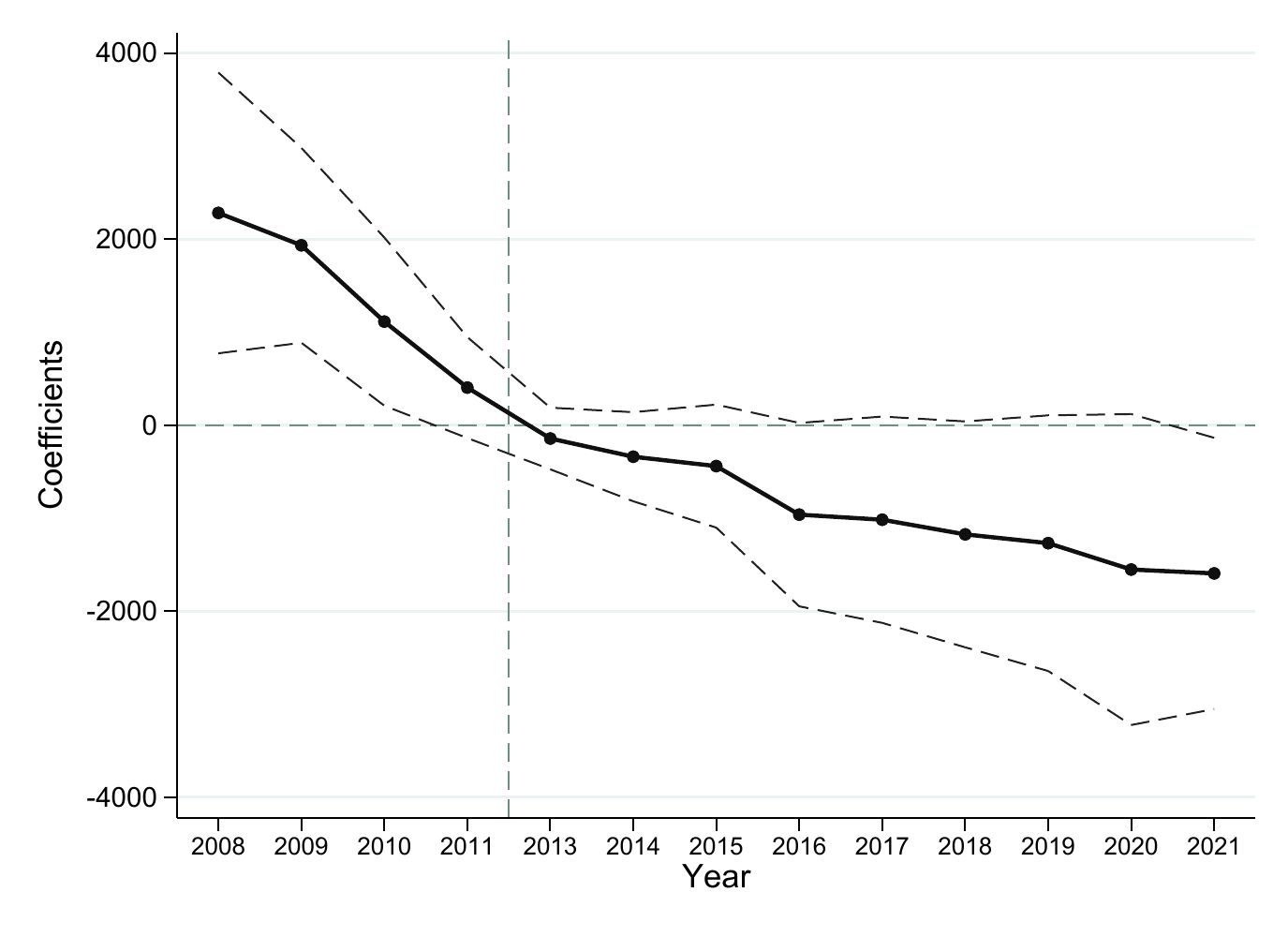}
    \end{minipage}%
    }%
    \caption{Dynamic Real Impact}

  \label{real}
  \end{figure}

According to the dynamic effects in Figure \ref{real} on page \pageref{real}, the short-term boost in both absolute and relative market share increased after the shock year as prosperity after the disaster. Labors experienced a long-term cut down, given the rigidity in employment. Peaks of R\&D and non-business expenditure are temporary. The fixed assets are likely to lower permanently. The effects on the retained earnings are ambiguous, which is reasonable because of the complex purpose of holding the fund.

  \begin{table}[H]
    \centering
    \scalebox{0.7}{
    \begin{threeparttable}
    \caption{Real impact}
    \label{did4}
      \begin{tabular}{llllllll}
          \toprule
          & (13)          & (14)          & (15)         & (16)                                 & (17)          & (18)   & (19) \\ 
          & rsms         & sms         & llabor      & rdg         & retained earnings & non bz expend & fix assets net \\ \hline
          ATT                 & 3.239719*    & .988167*     & -.617087*** & .117564***  & 1106.0513***       & 7.918901**      & -1846.1261***    \\
          & (1.584453)   & (.460682)    & (.199405)   & (.033326)   & (260.47375)        & (3.379832)      & (340.25711)      \\
CF in billion         & .692879**    & .142163      &             &             & 677.78537***       & 4.196696*       &                  \\
          & (.269379)    & (.121023)    &             &             & (45.462206)        & (2.25732)       &                  \\
lassets             & 5.950934***  & 1.053612***  &             &             & 647.46647***       & 7.051567***     &                  \\
          & (.777166)    & (.202867)    &             &             & (137.27222)        & (1.68046)       &                  \\
          non-tradable shares & -4.238499**  & -1.126026**  &             &             &                    &                 &                  \\
          & (1.635907)   & (.463971)    &             &             &                    &                 &                  \\
Employee in thousand  & .256584**    & .096394**    &             &             &                    &                 &                  \\
          & (.094466)    & (.035749)    &             &             &                    &                 &                  \\
CA in billion         & .034777**    & .007899*     &             &             & 121.77661***       & .583393***      &                  \\
          & (.011991)    & (.003722)    &             &             & (3.928001)         & (.080231)       &                  \\
lsalesvalue         & -9.745469*** & -2.533527*** &             &             &                    &                 &                  \\
          & (1.78227)    & (.817161)    &             &             &                    &                 &                  \\
gov                 &              &              & -.387648**  &             &                    &                 &                  \\
          &              &              & (.152453)   &             &                    &                 &                  \\
toptenholdersrate   &              &              & -.012523**  &             &                    &                 &                  \\
          &              &              & (.005647)   &             &                    &                 &                  \\
roa                 &              &              & -1.337148** &             & 3100.8001**        & -89.899363**    &                  \\
          &              &              & (.594029)   &             & (1362.8772)        & (31.001362)     &                  \\
sg                  &              &              &             & .68788***   &                    &                 &                  \\
          &              &              &             & (.0851)     &                    &                 &                  \\
L.net profit       &              &              &             & .000036***  &                    &                 &                  \\
          &              &              &             & (8.000e-06) &                    &                 &                  \\
non bz expend     &              &              &             & -.001549**  &                    &                 &                  \\
          &              &              &             & (.000579)   &                    &                 &                  \\
pg                  &              &              &             &             & -158.01113***      & 2.5847**        &                  \\
          &              &              &             &             & (55.308303)        & (1.128299)      &                  \\
ltangi              &              &              &             &             &                    &                 & 1384.223***      \\
          &              &              &             &             &                    &                 & (171.21089)      \\
sales               &              &              &             &             &                    &                 & .196545***       \\
          &              &              &             &             &                    &                 & (.010736)        \\
cons              & 85.082807*** & 25.47608**   & 8.097084*** & .084089***  & -5101.6272***      & -47.447384***   & -8873.3892***    \\
          & (21.285551)  & (9.612502)   & (.344856)   & (.007332)   & (1150.8944)        & (14.380565)     & (1356.2678)      \\
Observations        & 2713         & 2713         & 2713        & 1816        & 2109               & 2153            & 2850             \\
R-squared           & .857796      & .867778      & .807862     & .212055     & .873505            & .354141         & .836074         \\
          \bottomrule 
      \end{tabular}
      \begin{tablenotes}
        \footnotesize
            \item[]
            Standard errors are in parentheses. \\
            $ ^{***} p<.01, ^{**} p<.05, ^* p <.1 $ 
        \end{tablenotes}           
    \end{threeparttable}}   
  \end{table}

\section{Robust Tests}
\subsection{Placebo test}
This part applies the placebo test to emphasize the robustness of the Difference-in-difference effect of the basic model in Section 3. The placebo test is conducted through the following steps. First, resample the data by public firms, and assign a counterfactual event year for each firm's time series. On top of that, separate 150 of 328 total firms as treated groups randomly. Repeat the process 500 times to derive the average treatment effects on the treated.

\begin{figure}[H] 
      \centering 
      \includegraphics[width=0.7\textwidth]{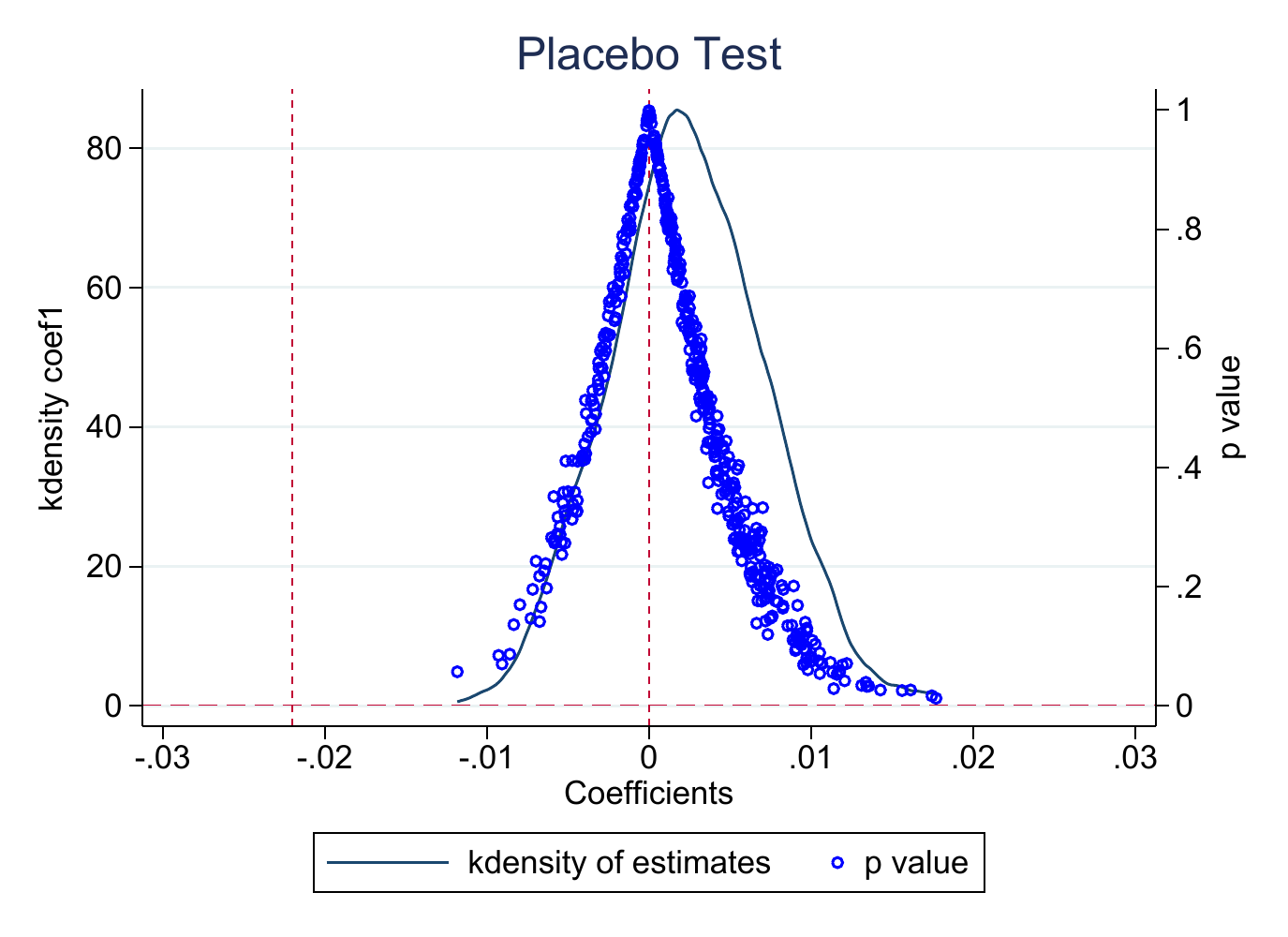} 
      \caption{Placebo Test} 
      \label{plb1}
    \end{figure}

As Figure \ref{plb1} on page \pageref{plb1} shows, the distribution of the estimates follows the normal distribution. The p-value of estimates is variable enough to reject the hypothesis that placebo counterfactual results are significant as the original group. The coefficients vary around x=0, while the previous estimation of causal effects is at x=-0.02. The placebo test strengthens the robustness of the baseline model's causal analysis. 

\subsection{PSM Test}
Rather than use some specific province data as the control group, in the PSM test, the whole national-wide data except Sichuan province are pooled to measure the propensity score.
Maximum likelihood estimation is applied to choose the covariates to accomplish sensitivity analysis. Nearest neighbor matching is used in PSM. Considering plenty of firms are available to be selected as the matched ones, the 20 nearest firms are matched to alleviate the fluctuations and satisfy the on-support hypothesis. Table \ref{TPSM1} in page \pageref{TPSM1} is the propensity score results in the firm level based pooled in time. T-stat does not consider the difference in time.

\begin{table}[H]
  \centering
  \begin{threeparttable}
  \caption{Propensity Score Results}
\label{TPSM1}
\begin{tabular}{lllllll}
\toprule
Variable & Sample     & Treated    & Controls    & Difference & S.E.       & T-stat \\ \hline
cfs     & Unmatched  & .102842192 & .096144639  & .006697553 & .006205271 & 1.08   \\
   &        ATT    & .102321667  & .095960611   &.006361056  & .006768436    & 0.94    \\
  &    ATU        &  .095976627  &  .094986798 &  -.000989829    &          &   \\
  &    ATE        &         &                &        -.00356335   &             & \\
\bottomrule
\end{tabular}
\end{threeparttable}       
\end{table}

\subsubsection{Balance Assumption Test}
The Balance test is applied to experiment whether there is a significant difference in the value of covariable between the two groups after matching. According to Table \ref{TPSM2} in page \pageref{TPSM2}, the \%bias of matched group is smaller than the unmatched data, far below 10\%. The hypothesis that there is no difference between the two groups for this covariable cannot be rejected. Besides, matched data are not distinguished under the examination of the t-test. To sum up, the difference is not obvious for the matched data, indicating that the matching effect is appropriate to be applied in DID regression. 

\begin{figure}[H] 
\centering 
\includegraphics[width=0.7\textwidth]{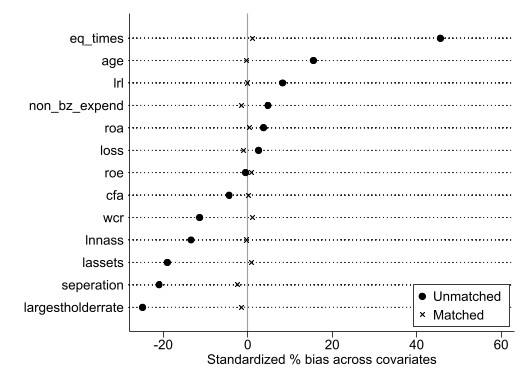} 
\caption{Bias under PSM} 
\label{psm1} 
\end{figure}

\begin{figure}[H]
\centering
\begin{minipage}[t]{0.48\textwidth}
\centering
\includegraphics[width=6cm]{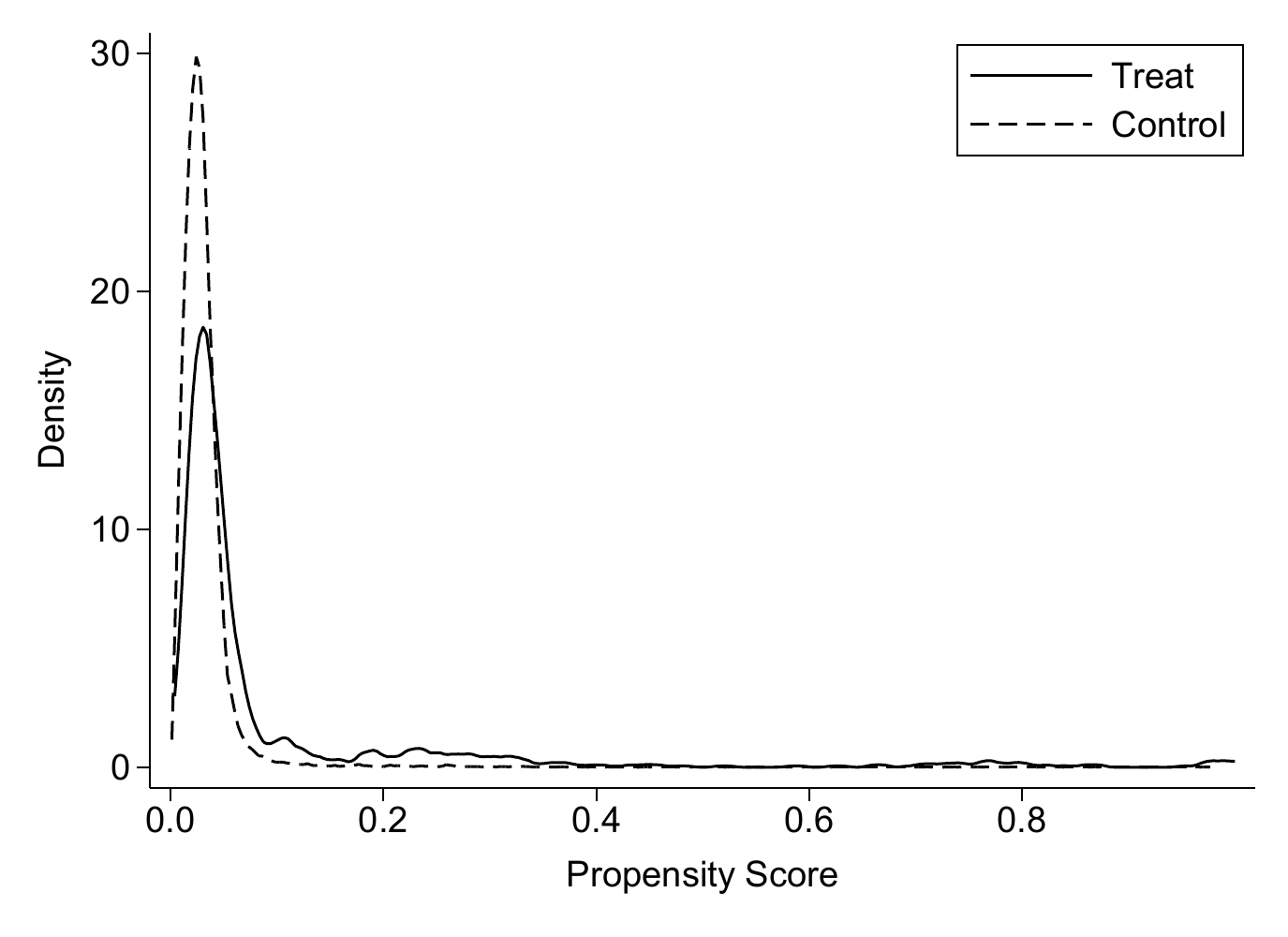}
\caption{Before Matching}
\label{psm2-1}
\end{minipage}
\begin{minipage}[t]{0.48\textwidth}
\centering
\includegraphics[width=6cm]{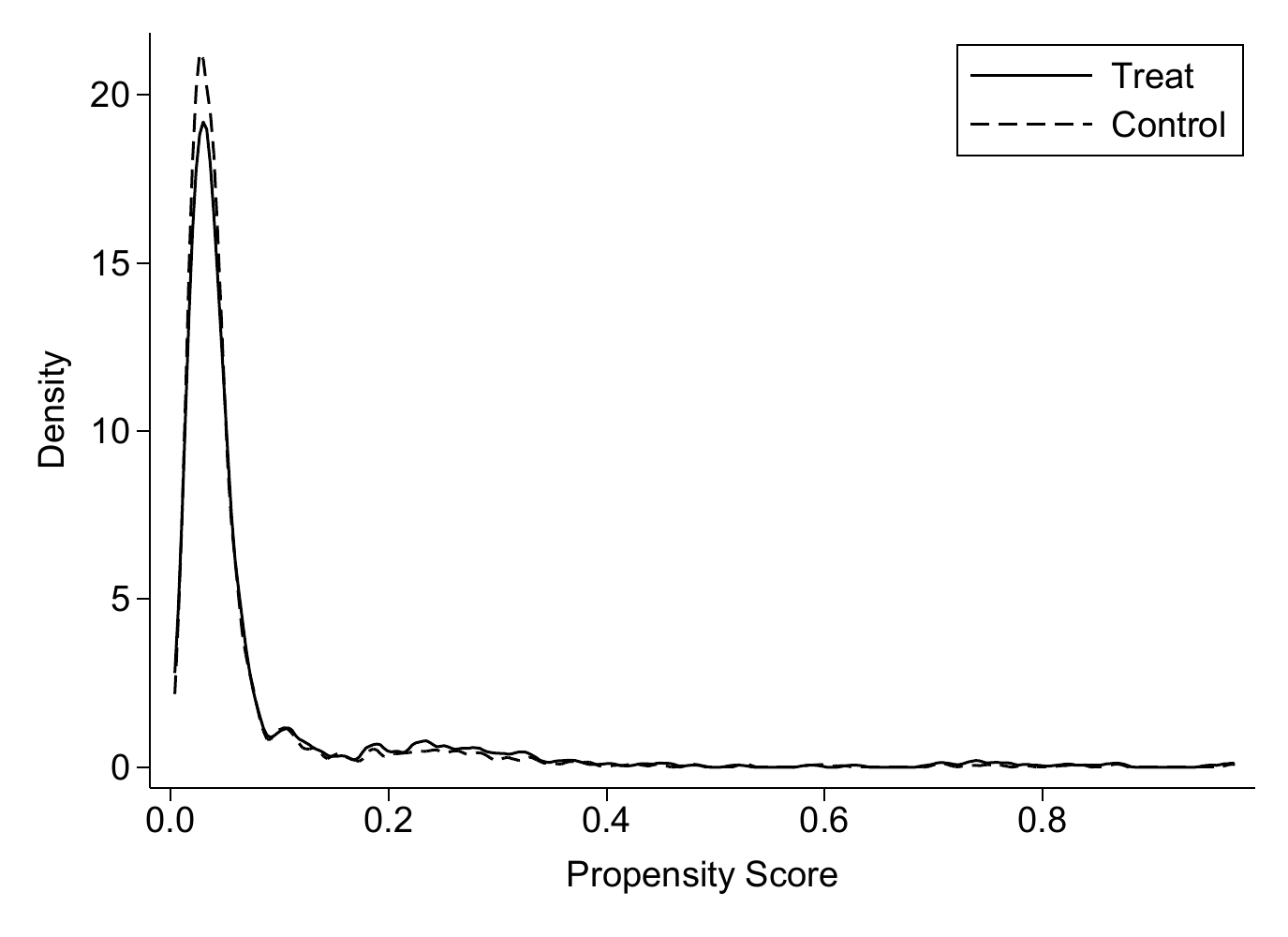}
\caption{After Matching}
\label{psm2-2}
\end{minipage}
\end{figure}

Table \ref{TPSM3} in page \pageref{TPSM3} illustrates the logit results after and before matching. The pseudo-R square in the regression results after matching becomes significantly smaller, indicating that there was little difference in the values of all covariables between the two groups after matching. Thus, the match process cannot explain the changes in the dependent variable in logit regression. Figure \ref{psm1} in page \pageref{psm1} is consistent with the above results to prove that there is no systematic deviation in the values of covariates between the two groups. The k-density curve in  Figure \ref{psm2-1} and \ref{psm2-2} supports the argument that the kernel density curve is closer after matching.

\begin{table}[H]
  \centering
  \begin{threeparttable}
  \caption{Balancing Assumption Test}
\label{TPSM2}
\begin{tabular}{lllllllll}
\toprule

 & Unmatched         &\multicolumn{2}{c}{Mean}     & \%reduct     &    &\multicolumn{2}{c}{t-test}        &  V(T)/ V(C)          \\
Variable          & Matched & Treated & Control & \%bias   & bias  & t     & p $ >$t    & \\ \hline
seperation        & U       & 3.1372        & 4.5567  & -21.0 &  & -5.82 & 0.000 & 0.74*  \\
& M                 & 3.1571  & 3.2759          & -1.8    & 91.6 & -0.41 & 0.682 & 1.07   \\
eq times          & U       & .23705          & .01341  & 45.6  &  &  38.13 & 0.000 & 30.92* \\
& M                 & .19466  & .18565           & 1.8     & 96.0  & 0.41  & 0.685 & 1.08   \\
age               & U       & 2.9754          & 2.9255  & 15.5  & & 4.64  & 0.000 & 1.04   \\
& M                 & 2.973  & 2.9771          & -1.3    & 91.9  & -0.28 & 0.782 & 1.20*  \\
lnnass            & U       & 7.5871       & 7.7488   & -13.4 & & -4.09 & 0.000 & 1.13   \\
& M                 & 7.5816  & 7.5906     &-0.7 &   94.5   & -0.16 & 0.875  & 1.09   \\
largestholderrate & U       & 31.229          & 34.793  & -24.9 &  & -7.14 & 0.000 & 0.87*  \\
& M                 & 31.221   &31.492   &  -1.9   & 92.4 & -0.41 & 0.679 &  0.98 \\
lassets           & U       & 8.2027  & 8.3662   & -12.4 & & 3.69 & 0.000 & 1.01   \\
& M                 & 8.1962  & 8.2185 &    -1.7   & 86.3 &  -0.36 & 0.716 &  1.02\\

non bz expend     & U       & 17.84  &  15.573    &   4.8   & & 1.46  & 0.144 & 1.13   \\
& M                 &  17.54  &   18.12   &   -1.2  &   74.4 &   -0.24&   0.811 &   0.80* \\
lrl               & U       & .0886           & .07307  & 8.3   &  & 2.58  & 0.010 & 1.25*  \\
& M                 & .08831  & .08942          & -0.6    & 92.8&  -0.13 & 0.900 &  1.26* \\
loss              & U       & .10915          & .10126  & 2.6   &  & 0.77  & 0.439 & .      \\
& M                 &  .11012   &.11613  &   -2.0  &  23.6 &  -0.40 & 0.688  & .      \\
cfa               & U       & .046            & .04901  & -4.4  & & -1.28 & 0.199 & 0.91   \\
& M                  &.04566    &.04503    &   0.9    & 79.1  &  0.19 &  0.847 &  0.85*   \\
roa               & U       & .04198          & .03953  & 3.7   & & 1.12  & 0.263 & 1.04   \\
& M                 &  .04181  &  .04163   &    0.3   &  92.7 &    0.06 &  0.954 &   1.01  \\
roe               & U       & .05702          & .05775  & -0.5  & & -0.15 & 0.880 & 1.16*  \\
& M                 & .05669   & .05535    &   0.9  &  -91.1 &   0.18 &  0.856 &   0.95 \\
\bottomrule
\end{tabular}
\begin{tablenotes}
\footnotesize
\item[]
$ ^{*}$ if variance ratio outside [0.88; 1.14] for U and [0.88; 1.14] for M\\
\end{tablenotes}            
\end{threeparttable}       
\end{table}

\begin{table}[H]
  \centering
  \begin{threeparttable}
  \caption{Balancing Assumption Test}
\label{TPSM3}
\begin{tabular}{lllllllll}
\toprule

Sample    & Ps R2 & LR chi2 & p\textgreater{}chi2 & MeanBias & MedBias & B     & R      & \%Var \\
Unmatched & 0.082 & 653.00  & 0.000               & 12.2     & 8.2     & 56.1* & 12.26* & 50    \\
Matched   & 0.001 & 1.27    & 1.000               & 1.3      & 1.3     & 5.3   & 1.19   & 42 \\ \bottomrule
\end{tabular}
\begin{tablenotes}
\footnotesize
\item[]
if B$>$25\%, R outside [0.5; 2]\\
\end{tablenotes}            
\end{threeparttable}       
\end{table}

%reference:
%https://zhuanlan.zhihu.com/p/90193649
%https://www.lianxh.cn/news/897d7077ccfed.html
%https://blog.csdn.net/arlionn/article/details/90108138
%https://zhuanlan.zhihu.com/p/392443329
%important : https://www.lianxh.cn/news/9a4ddf5ddaac8.html?luicode=10000011&lfid=1076031999313777&u=https%3A%2F%2Fwww.lianxh.cn%2Fnews%2F9a4ddf5ddaac8.html

\subsubsection{PSM regression}
The regression after propensity score matching is divided into (20) to (22). (21) is the original regression as a benchmark. (21) implements the sample where weight is not 0, i.e., successful matching. (22) implements the sample by computing the frequency of the firms emerging to measure the importance of specific firms during matching. The regression by weight behaves best. The causal effects are negative in all three regressions. The magnitude in (21) is slightly larger in absolute value than (1) -(4). (20) is potentially underestimated concerning the different groups of control variables. Besides, industry fixed effects are included because the matching sample is all public firms in China. The PSM estimation sustains that selection bias does not affect the robustness of the baseline regression results.

\begin{table}[H]
  \centering
  \begin{threeparttable}
  \caption{PSM-DID}
\label{psmdid}
\begin{tabular}{llll}
\toprule
& (20)         & (21)         & (22)         \\
&   Original      & By if weight        & Weight by frequency      \\ \hline
ATT               & -.009946    & -.051959**  & -.04189     \\
                  & (.008502)   & (.021918)   & (.026758)   \\
Observations      & 35183       & 10962       & 537578      \\
R-squared         & .777805     & .849063     & .909566    \\
\bottomrule
\end{tabular}
\begin{tablenotes}
\footnotesize
\item[]
Standard errors are in parentheses.\\
$ ^{***} p<.01, ^{**} p<.05, ^* p <.1$\\
\end{tablenotes}           
\end{threeparttable}       
\end{table}

\section{PSM-DID on multiple strong earthquakes}
Given the support of previous results, the probable effects of a specific moderate and strong earthquake are unfolding. This section generalizes the question to focus on multiple earthquakes at the different time rather than the greatest one, the Lushan Earthquake. The assumed probable influence range is shrunk to the city, rather than the province level. 

The spatial allocation of the public firms' office addresses and the strong earthquakes are merged by the city. The city division is according to the administration code. Semester data are collected and applied, rather than annual data, to prevent the overlap of two earthquakes that happened consecutively, though the variables are the same. The time spreads from 2008Q1 to 2021Q4. The table \ref{eq2} on page \pageref{eq2} describes the frequency of earthquakes with a magnitude higher than 4.5, as well as what happened in the geographical locations of some public firms. Thus, not all drastic earthquakes are included, such as the Lushan Earthquake, because there are no public firms in the city-level region of Lushan. The merge methods are precise to restrict the potential effect of strong earthquakes in a reasonable range, which benefits the analysis of the direct impact of discrete multiple natural hazards that happened at different time on corresponding firms.

\begin{table}
  \centering
  \scalebox{0.8}{
\begin{threeparttable}         
\caption{Strong Earthquakes Frequency}
\begin{tabular}{llllllllllllllllllllllllllllllllll}
\toprule
\multicolumn{2}{l}{M}         & \multicolumn{2}{l}{0}       & \multicolumn{2}{l}{4.5} & \multicolumn{2}{l}{4.6} & \multicolumn{2}{l}{4.7} & \multicolumn{2}{l}{4.8} & \multicolumn{2}{l}{4.9} & \multicolumn{2}{l}{5}  & \multicolumn{2}{l}{5.1} & \multicolumn{2}{l}{5.2} & \multicolumn{2}{l}{5.3} & \multicolumn{2}{l}{5.4} & \multicolumn{2}{l}{5.6} & \multicolumn{2}{l}{6} & \multicolumn{2}{l}{6.1} & \multicolumn{2}{l}{6.4} & \multicolumn{2}{l}{Total}   \\ \hline
\multicolumn{2}{l}{Frequency} & \multicolumn{2}{l}{135,364} & \multicolumn{2}{l}{6}   & \multicolumn{2}{l}{63}  & \multicolumn{2}{l}{7}   & \multicolumn{2}{l}{13}  & \multicolumn{2}{l}{11}  & \multicolumn{2}{l}{10} & \multicolumn{2}{l}{35}  & \multicolumn{2}{l}{5}   & \multicolumn{2}{l}{2}   & \multicolumn{2}{l}{1}   & \multicolumn{2}{l}{1}   & \multicolumn{2}{l}{2} & \multicolumn{2}{l}{1}   & \multicolumn{2}{l}{2}   & \multicolumn{2}{l}{135,523}\\
 \bottomrule
\label{eq2}
\end{tabular}       
\end{threeparttable} }       
\end{table}

Strong earthquakes are unusual, so the estimation follows the PSM-DID in section 5 as a benchmark. The 5 nearest neighbor firms are matched to alleviate the fluctuations and satisfy the on-support hypothesis in estimating the propensity score. Multi-way fixed effects are estimated by clustering in firms, cities, time, and industry. The PSM-DID estimation results on cash flow in the table \ref{psmevt} on page \pageref{psmevt} illustrates that the cash flow would still be struck by moderate and strong earthquakes, though we relax the conditions on the magnitude is great. As a contrast, the estimated causal effects in the model(23)-(25) are relatively higher than Table \ref{psmdid}. It is reasonable because of the variation in variable selection for estimating propensity score. The negative shocks are more significant considering the refined definition in the range of affected regions by the earthquakes. The semester data is also more sensitive than annual statistics. 

\begin{equation}
  cfs=\alpha+X'_{i,t}\Gamma+ \tau post_{i,t}*Treat_{i,t}+\delta_t+\theta_i+ indus_i+\theta_i*indus_i + citycode_{i,t}+ \epsilon_{i,t}
  \end{equation}

\begin{figure}[H] 
  \centering 
  \includegraphics[width=0.7\textwidth]{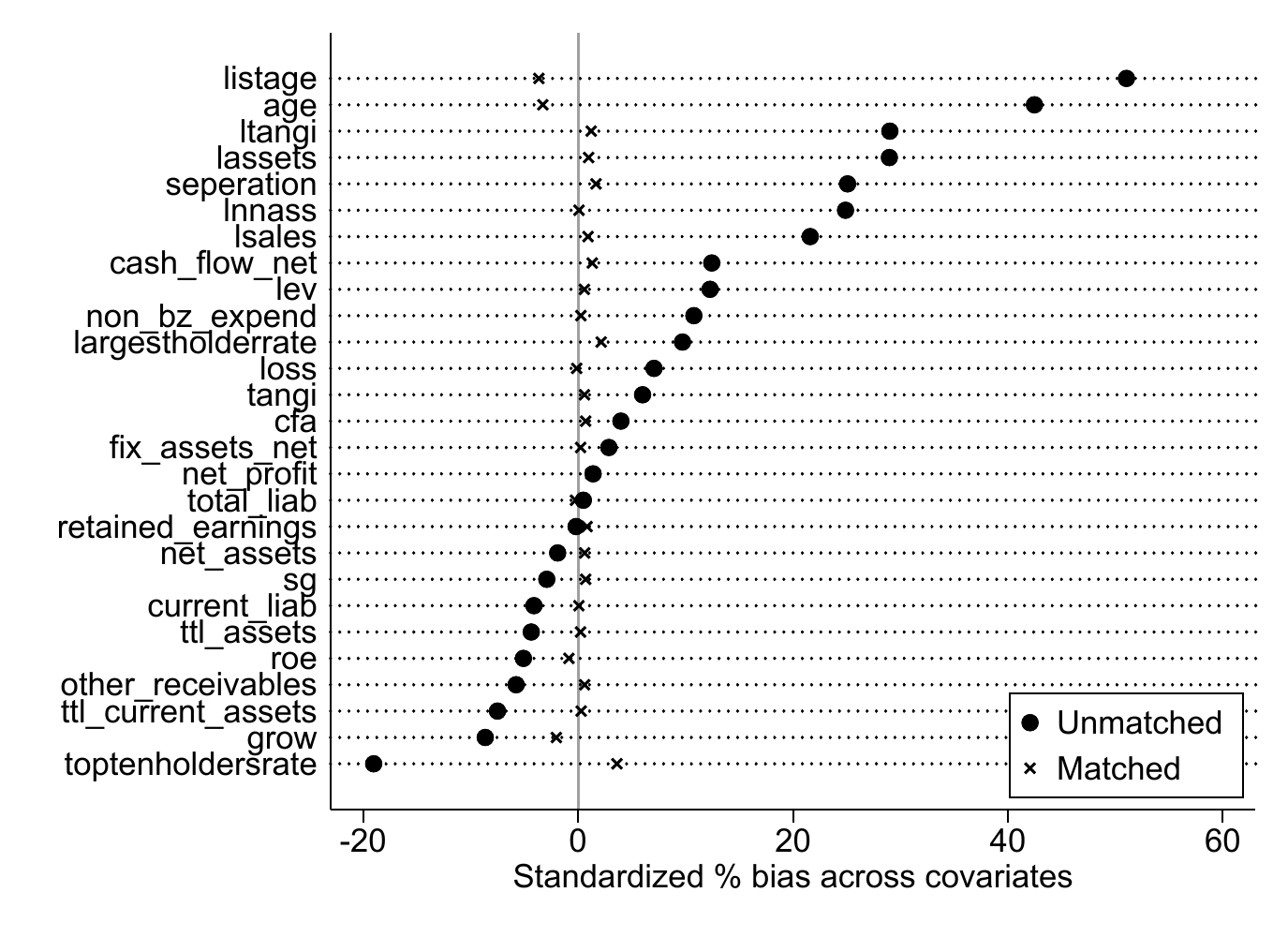} 
  \caption{Bias under PSM} 
  \label{psm3} 
  \end{figure}

  \begin{figure}[H]
  \centering
  \begin{minipage}[t]{0.48\textwidth}
  \centering
  \includegraphics[width=6cm]{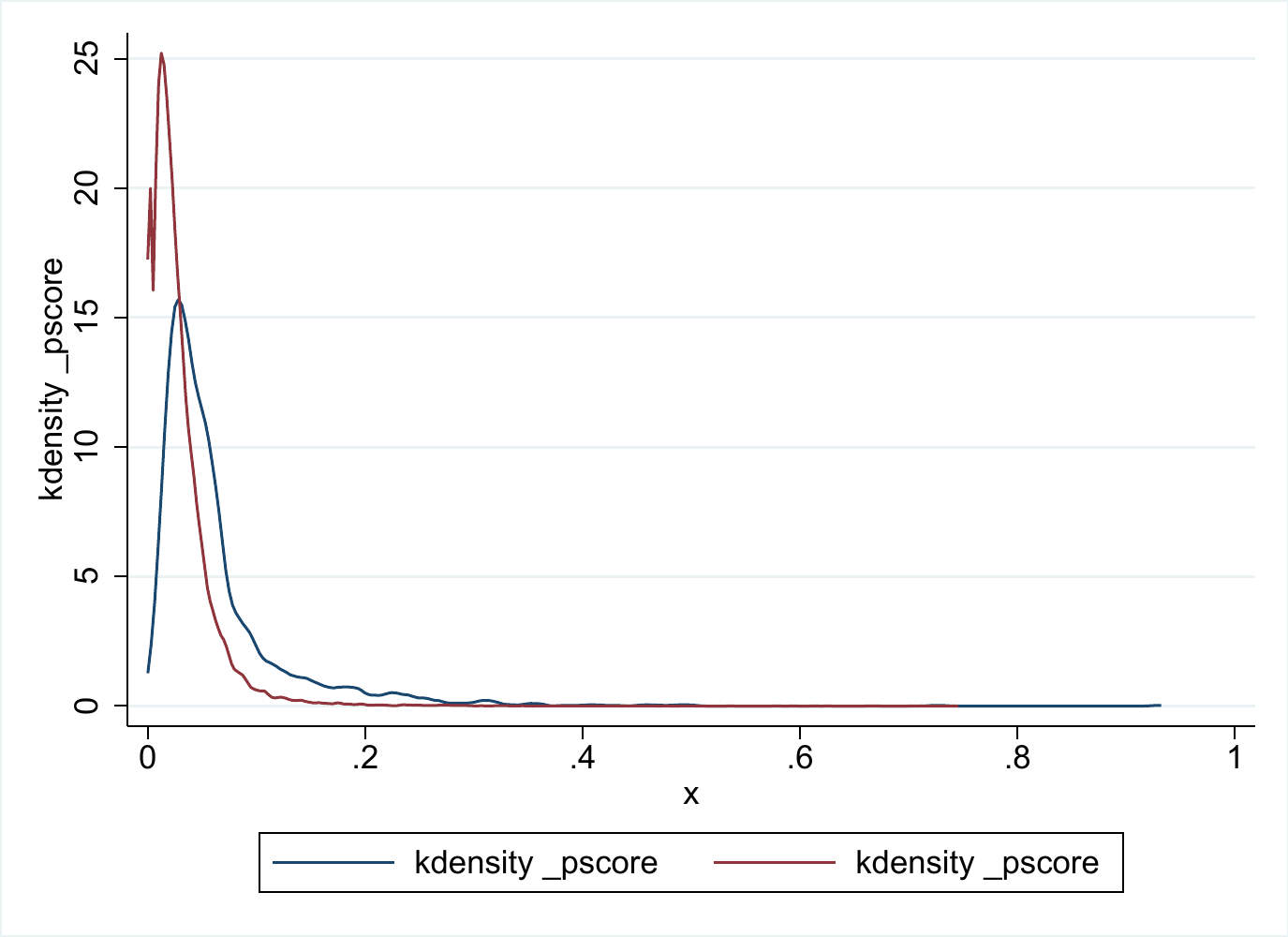}
  \caption{Before Matching}
  \label{psm4-1}
  \end{minipage}
  \begin{minipage}[t]{0.48\textwidth}
  \centering
  \includegraphics[width=6cm]{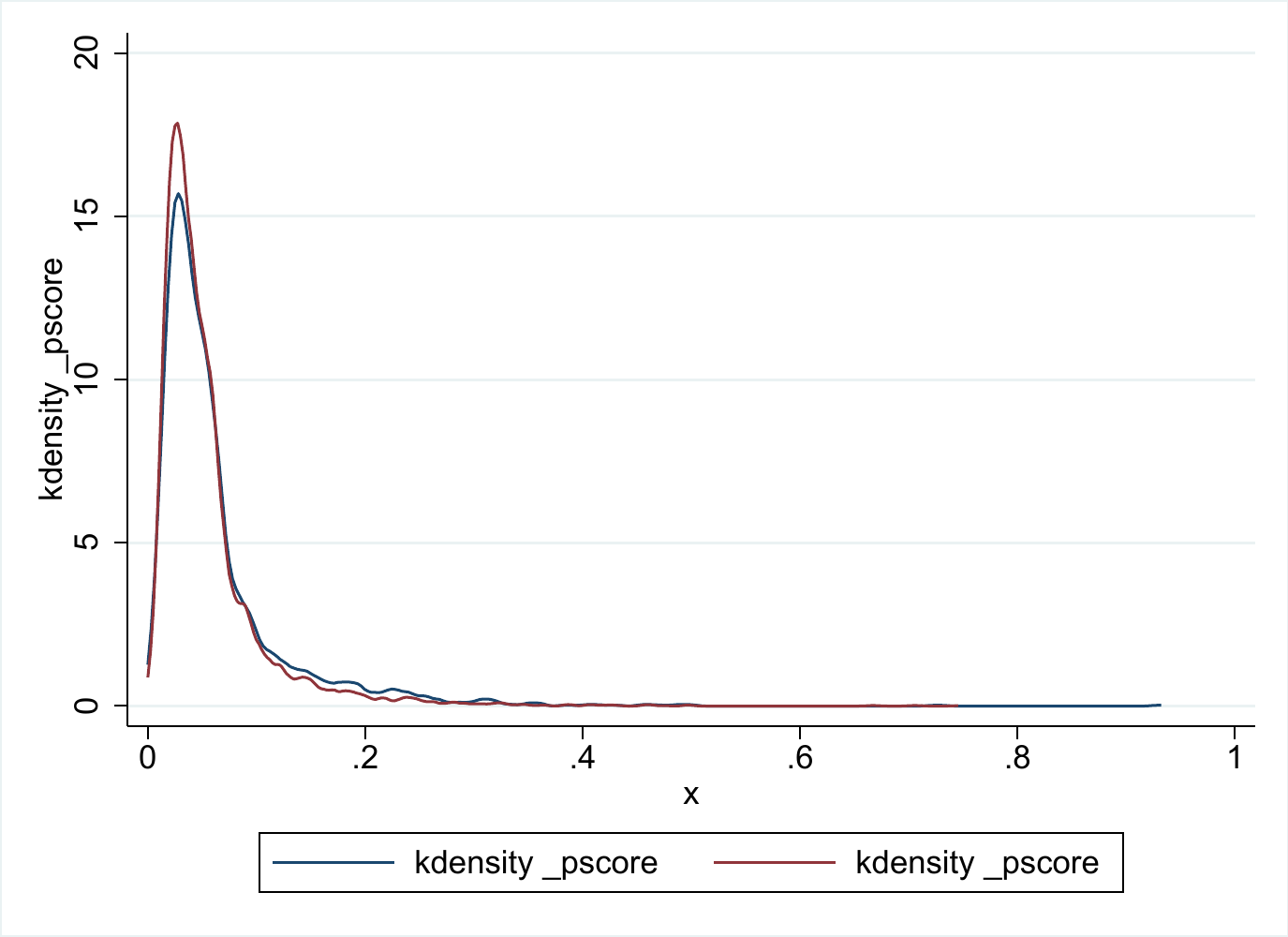}
  \caption{After Matching}
  \label{psm4-2}
  \end{minipage}
  \end{figure}

\begin{table}[H]
  \centering
  \begin{threeparttable}
  \caption{PSM-DID of Multiple Earthquakes}
\label{psmevt}
\begin{tabular}{llll}
\toprule
& (23)         & (24)         & (25)         \\
cfs&   Original      & By if weight        & Weight by frequency      \\ \hline
treat\_post  & -.298871**   & -.130422     & -.533093***  \\
& (.124849)   & (.084965)   & (.145471)   \\
\_cons       & 5.212762*** & 4.846632*** & 5.242664*** \\
& (.025268)   & (.00256)    & (.025437)   \\
Observations & 10722       & 72307       & 12410       \\
R-squared    & .750006     & .676723     & .678377    \\
\bottomrule
\end{tabular}
\begin{tablenotes}
\footnotesize
\item[]
Standard errors are in parentheses.\\
$ ^{***} p<.01, ^{**} p<.05, ^* p <.1$\\
\end{tablenotes}           
\end{threeparttable}       
\end{table}

\section{Event Study}

Event study is a statistical method to evaluate the impact of a particular event on firms. For each firm, event study could serve as the estimating method if they were once shocked by a sudden earthquake in a particular year. With the diff-in-diff method combined, the firms could be divided into a control group and a treated group on whether the city of the firms was shocked within the time frame. The earthquakes happened at different times, so the timelines of each firm are adjusted respectively according to the earthquake happen time. The advantage of semester data is that there is almost no overlapping between the time window of earthquakes. For example, firms may experience moderate earthquakes the other year. At least 4 leads' data should be reserved to ensure data integrity. Thus, the semester timeline benefits from retaining more valid data. The previous analysis also supports that the shock of earthquakes on sales and cashflows is more likely to recover in one year. This type of impact is temporary.

Event study on the firms' loss caused by multiple earthquakes that happened at different times is conducted according to the method in \cite{clarke2021implementing}. The regression is robust to heteroskedasticity by clustering in the semester, city, firm level, industry, and firm combined with industry, similar to PSM-DID design in the case of multiple earthquakes. The time range is from 2008Q1 to 2021Q4. The city code is a dominant variable to merge in spatial because the administrative division takes the economic development level and geographical proximity into account, such that the division is neither overlapping nor largely distinguishing of the adjacent area. 

\begin{figure}[H]
  \centering
\caption{ Firm loss following local strong earthquakes}
  \subfigure[Cash Flow/Sales ]{
  \begin{minipage}[t]{0.5\linewidth}
  \centering
  \includegraphics[width=3in]{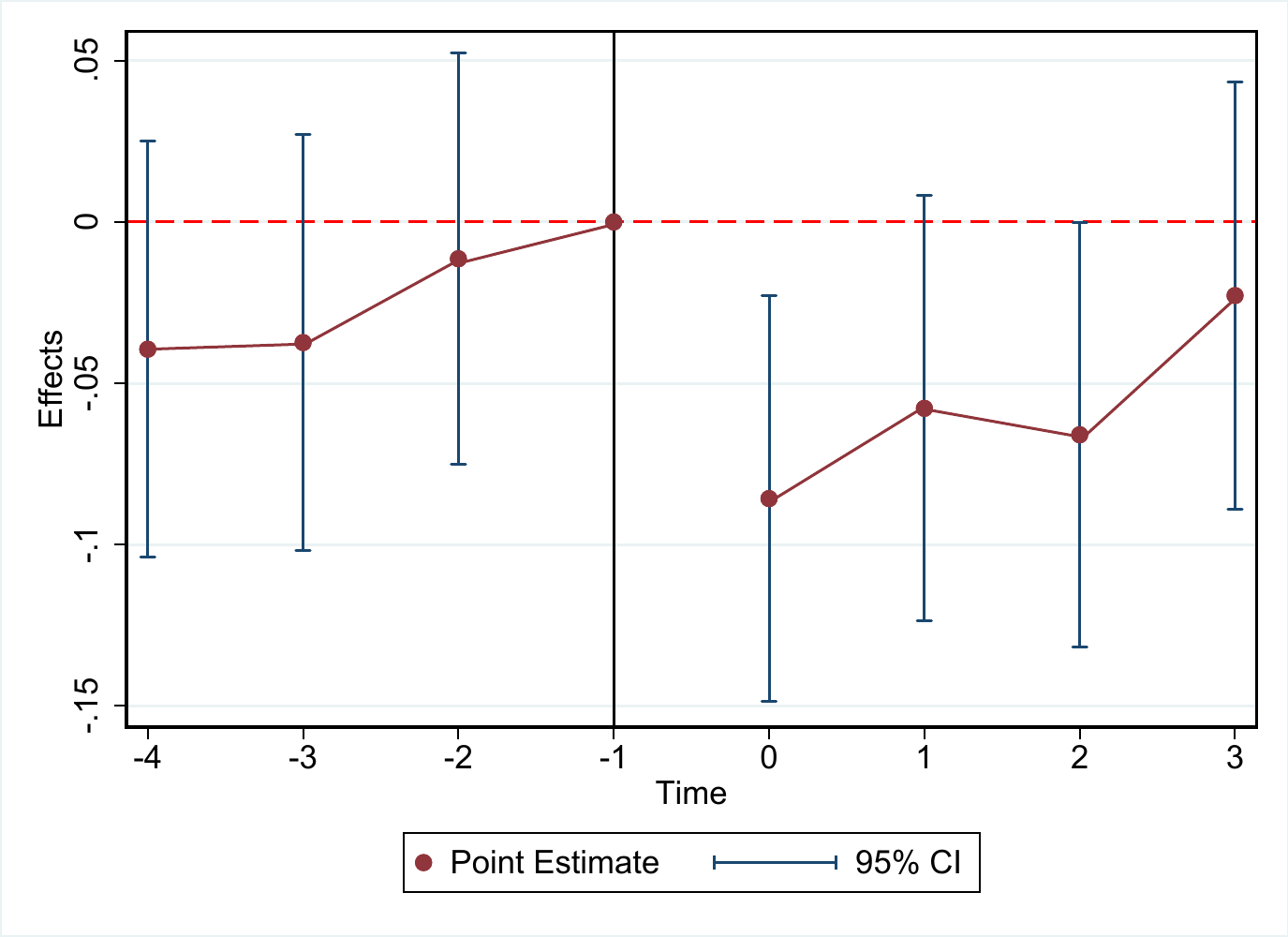}
  \end{minipage}%
  }%
  \subfigure[Sales]{
  \begin{minipage}[t]{0.5\linewidth}
  \centering
  \includegraphics[width=3in]{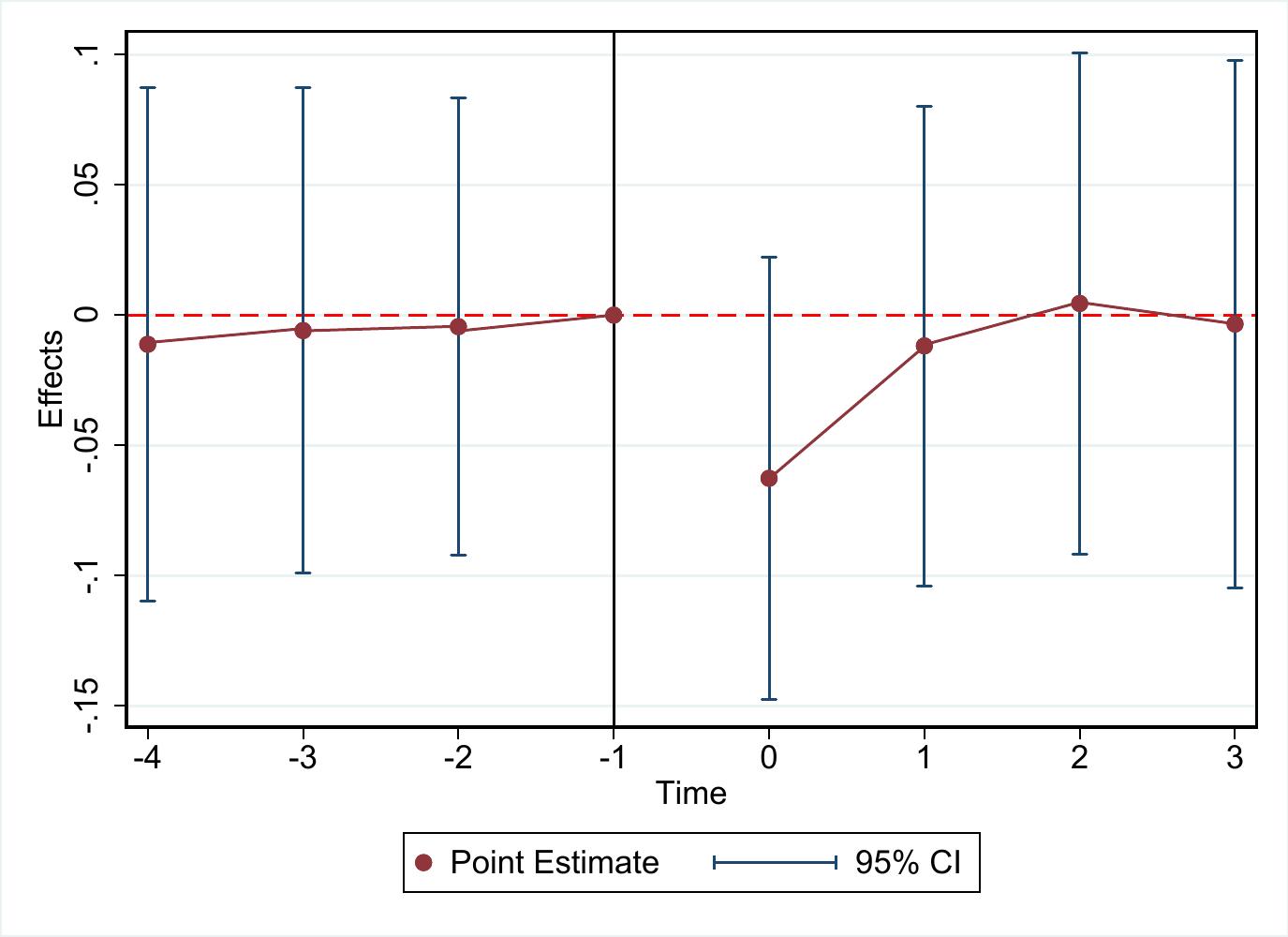}
  \end{minipage}%
  }%

  \label{eventfg}
  \end{figure}

Figure \ref{eventfg} on page \pageref{eventfg} distinctly demonstrates that the ratio of cash flow to firms dropped sharply in the following 3 semesters. Consistent with the great Lushan Earthquake, the shock on the ratio of cash flow to sales recovered in 3 subsequent semesters, i.e., one year. In contrast, sales have a higher recovery velocity. The shock is quite temporary. The rapid sales recovery could account for the insignificance of sales level in the Lushan Earthquake case because semester drop is hard to observe in annual reports. Considering the magnitude of a filtered earthquake is smaller than 6.5, the relatively slight adverse effects are reasonable. (26) and (27) are analogous to (5) and (6), respectively. The negative influence on CF/Sales is higher in (26) to favor the efficiency of the multiple earthquakes analysis. Although the magnitude of these earthquakes is non-comparable to the Great Lushan Earthquake, the impacts are more direct, so the firms' reactions are more rapid.

Furthermore, Table \ref{eventdd} on page \pageref{eventdd} measures the effect of the current period. Compared with Figure \ref{real} on page \pageref{real}, the cut-down is more significant in Figure \ref{eventfg}. Focusing on the firms shocked directly by a specific earthquake would enhance the estimation efficiency by reducing the noise. Such noise is intense in the case of the Lushan Earthquake because the firms located at outer strata from the epicenter would suffer less than the closer ones. 

\begin{align}
  cfs=  &\Sigma _{\tau=-4}^{3} \beta_{1,\tau}\times Treat_{i,t}\times if\_eq_{i,t-\tau}+ \Sigma _{\tau=-4}^{3} \beta_{2,\tau} \times Treat_{i,t}   +\Sigma _{\tau=-4}^{3}\beta_{3,\tau} \times  if\_eq_{i,t-\tau}  \notag\\
  &  + \alpha+X'_{i,t}\Gamma+  \delta_t+\theta_i+ indus_i+\theta_i \times indus_i + citycode_{i,t}+ \epsilon_{i,t}
\end{align}

\begin{table}[H]
  \centering
  \scalebox{0.8}{
  \begin{threeparttable}
  \caption{Event Study Results}
\label{eventdd}
\begin{tabular}{lll}
\toprule
& (26)         & (27)        \\
& cfs         & lsales     \\ \hline
net\_profit  & .000017***  &            \\
& (3.000e-06) &            \\
lsales       & .033449***  &            \\
& (.002568)   &            \\
grow         &             & .042505*** \\
&             & (.01354)   \\
lassets      &             & .686879*** \\
&             & (.037098)  \\
lnnass       &             & .159849*** \\
&             & (.036941)  \\
lev          &             & .401013*** \\
&             & (.093401)  \\
age          &             & -.185642** \\
&             & (.092136)  \\
lead4        & -.039433    & -.011188   \\
& (.032922)   & (.050181)  \\
lead3        & -.037367    & -.005921   \\
& (.032867)   & (.047517)  \\
lead2        & -.011396    & -.004427   \\
& (.032511)   & (.044722)  \\
lag0         & -.085714*** & -.062611   \\
& (.032114)   & (.043215)  \\
lag1         & -.057746*   & -.011803   \\
& (.033623)   & (.04695)   \\
lag2         & -.06594**   & .004478    \\
& (.033536)   & (.04901)   \\
lag3         & -.022797    & -.003421   \\
& (.033804)   & (.05158)   \\
\_cons       & -.134408*** & .460636    \\
& (.028993)   & (.287151)  \\
Observations & 6346        & 6098       \\
R-squared    & .061028     & .955567    \\
\bottomrule
\end{tabular}
\begin{tablenotes}
\footnotesize
\item[]
Standard errors are in parentheses.\\
$ ^{***} p<.01, ^{**} p<.05, ^* p <.1$\\
\end{tablenotes}           
\end{threeparttable} }
\end{table}

\section{Conclusion}

The Great Lushan Earthquake that happened in 2013 in China was introduced as a sudden shock. By comparing the average treatment effects between treated firms in Sichuan Province, and controlled firms in Henan and Hebei Province with the Difference-in-Difference method, the study initially demonstrates that the liquidity of cash flow is significantly negatively affected in the short-term by the sudden catastrophic shock of the Great Lushan Earthquake. Furthermore, the causal effects on cash flow are rather strong to drive the negative trend of relative ratios, such as cash flow to assets, and cash flow to sales. However, the sales and assets cannot derive the same, even weaker trend independently without the combined influence of cash flow cutting down. Profits and cash flow have strong correlation effects, while the existence of causal effects cannot be assured. The baseline model of the causal effects of earthquakes on cash flow scaled by sales is robust in both PSM regression and placebo tests.

Besides, the firms under the shadow of such natural disasters experienced the real impact on manufacturing and operation. The treated public firms suffered from the subsequent labor reduction than the controlled public firms in other provinces, which were not affected directly by the earthquake. Concerning the rigidity in employment, the drop in employment level has been nearly permanent in the following years until now. Another long-term payment of this unexpected shock is fixed assets, which is firmly connected with the property of the cohort composition in the treated and control group. The majority of companies are manufacturing enterprises, and the fixed cost of production is considerable. If ever interrupted, the vulnerability of the production continuity would surpass the expectations of risk control and bring firms excessive loss.

In the short term, firms increase non-business expenditure to relieve from the sudden shock. Firms have the tendency to increase retained earnings as resistance to potential following series risk. The recovery from such a disaster would benefit the companies in the following years. The manufacturing firms may face growth in demand considering post-disaster reconstruction. Social assistance would also positively promote firms to raise research and development investment. The growth rate of R\&D increases over the basis of the preceding year significantly. Furthermore, the treated firms were boosted after the earthquake in improving the market competitiveness. Both the relative market share and the absolute market share elevate to some degree. 

Empirical results of the impact of multiple strong earthquakes serve as supplementary. 159 earthquakes with magnitudes higher than 4.5 are retained because they happened in the city where public firms are located, from 2008Q1 to 2021Q4. PSM-DID and event study methods are implemented to investigate the general effects of a particular earthquake on local public firms. Consistent with the Lushan Earthquake, the ratio of cash flow to sales dropped drastically and recovered in 3 subsequent semesters. The shock on sales was more transitory only in the current semester, accounting for the insignificance of sales in the Lushan Earthquake case. 

To sum up, unexpected natural hazards like earthquakes have complicated and comprehensive impacts on firms' development. The long-term negative effects of production factors and short-term ambiguous effects interact with the firms. The comprehensive causal effects are valid in both single great earthquakes and multiple moderate and strong earthquakes. The frequency of such severe catastrophes is rather low, so such a possibility is less likely to be the decisive factor in selecting the factories and offices' locations. The paper provides the basic empirical evidence about the comprehensive causal effects of a single great earthquake and integrated multiple strong earthquakes on the firms. Future studies could focus on the risk propagation of multiple earthquake shocks through the firms' financial network and further the network spillover effects to explain the risk tolerance and stability of firm operation.

\nocite{*}
\bibliography{paper} 
\end{document}